\documentclass[useAMS,usenatbib,a4paper]{mn2e}

\input psfig.sty
\usepackage{xspace}
\usepackage{graphicx}
\usepackage{amssymb}
\usepackage{amsmath}

\graphicspath{{./}{sw-system-figs/}{sw-system-figs/gallery/}}

\newcommand{\ie}{{\it i.e.\ }}
\newcommand{\etc}{{\it etc.\ }}

\newcommand{\be}{\begin{equation}}
\newcommand{\ee}{\end{equation}}
\newcommand{\bea}{\begin{eqnarray}}
\newcommand{\eea}{\end{eqnarray}}

\def\Sref#1{Section~\ref{#1}\xspace}
\def\Fref#1{Figure~\ref{#1}\xspace}
\def\Tref#1{Table~\ref{#1}\xspace}
\def\Eref#1{Equation~\ref{#1}\xspace}
\def\Aref#1{Appendix~\ref{#1}\xspace}

\newcommand{\PaperTwo}{Paper~II\xspace}

\newcommand{\StageOne}{Stage~1\xspace}
\newcommand{\StageTwo}{Stage~2\xspace}

\newcommand{\kms}{\ifmmode  \,\rm km\,s^{-1} \else $\,\rm km\,s^{-1}  $
\fi }
\newcommand{\kpc}{\ifmmode  {\rm kpc}  \else ${\rm  kpc}$ \fi  }
\newcommand{\pc}{\ifmmode  {\rm pc}  \else ${\rm pc}$ \fi  }
\newcommand{\Msun}{\ifmmode {\rm M_{\odot}} \else ${\rm M_{\odot}}$ \fi}
\newcommand{\Zsun}{\ifmmode {\rm Z_{\odot}} \else ${\rm Z_{\odot}}$ \fi}
\newcommand{\yr}{\ifmmode yr^{-1} \else $yr^{-1}$ \fi}
\newcommand{\hMsun}{\ifmmode h^{-1}\,\rm M_{\odot} \else $h^{-1}\,\rm
M_{\odot}$ \fi}

\def\sw{{\small\sc Space\,Warps}\xspace}
\def\SW{{\sc Space\,Warps}\xspace}
\def\Talk{{\small\sc Talk}\xspace}

\def\cfhtls{{CFHTLS}\xspace}

\def\gravlens{{\sc gravlens}\xspace}

\def\fitsjs{{\sc fitsjs}\xspace}
\def\humvi{{\sc HumVI}\xspace}

\def\RF{{\sc RingFinder}\xspace}
\def\PH{{\sc Planet\,Hunters}\xspace}
\def\GZ{{\sc Galaxy\,Zoo}\xspace}

\def\pr{{\rm Pr}}
\def\data{{\mathbf{d}}}

\def\training{{\mathbf{d}^{\rm t}}}
\def\trainingk{{\mathbf{d}^{\rm t}_k}}

\def\LENS{{\rm LENS}}
\def\saidLENS{{\rm ``LENS"}}
\def\NOT{{\rm NOT}}
\def\saidNOT{{\rm ``NOT"}}
\def\CM{\mathcal{M}}
\def\PL{\CM_{LL}}
\def\PD{\CM_{NN}}

\def\effort{N_{\rm C}}
\def\thiseffort{N_{{\rm C},k}}
\def\experience{N_{\rm T}}
\def\skill{{\langle \Delta I \rangle_{0.5}}}
\def\contribution{\skill^{\rm total}}
\def\information{\Delta I}
\def\Ns{J} 
\def\Nv{K} 
\def\Ncands{N_{\rm det}} 

\usepackage[usenames]{color}

\def\oxford{Dept.\ of Physics, University of Oxford, Keble Road, Oxford, OX1 3RH, UK}
\def\oxfordeng{Dept.\ of Engineering Science, University of Oxford, Parks Road, Oxford, OX1 3PJ, UK}
\def\kipac{Kavli Institute for Particle Astrophysics and Cosmology, Stanford University, 452 Lomita Mall, Stanford, CA 94035, USA}
\def\ipmu{Kavli IPMU (WPI), UTIAS, The University of Tokyo, Kashiwa, Chiba 277-8583, Japan}
\def\zooniverse{Zooniverse, c/o Astrophysics Department, University of Oxford, Oxford OX1 3RH, UK}
\def\adler{Adler Planetarium, Chicago, IL, USA}

\def\zurich{Department of Physics, University of Zurich, Winterthurerstrasse 190, 8057 Zurich, Switzerland}

\def\icg{Institute of Cosmology and Gravitation, University of Portsmouth, Dennis Sciama Building, Portsmouth P01 3FX, UK}

\def\pjmemail{\tt pjm@slac.stanford.edu}

\title[\SW]
{\SW: I. Crowd-sourcing the Discovery of Gravitational Lenses}

\author[Marshall et al.]{%

   \newauthor{%
    Philip~J.~Marshall,$^{1,2}$\thanks{\pjmemail}
    Aprajita~Verma,$^{2}$
    Anupreeta~More,$^{3}$
    Christopher~P.~Davis,$^{1}$
    }
    \newauthor{%
    Surhud~More,$^{3}$
    Amit~Kapadia,$^{4}$
    Michael~Parrish,$^{4}$
    Chris~Snyder,$^{4}$
    }
   \newauthor{%
    Julianne~Wilcox,$^{5}$
    Elisabeth~Baeten,$^{5}$
    Christine~Macmillan,$^{5}$
    Claude~Cornen,$^{5}$
    }
   \newauthor{%
    Michael~Baumer,$^{1}$
    Edwin~Simpson,$^{6}$
    Chris~J.~Lintott,$^{2}$
    David~Miller,$^{4}$
    }
   \newauthor{%
    Edward~Paget,$^{4}$
    Robert~Simpson,$^{2}$
    Arfon~M.~Smith,$^{4}$
    Rafael~K\"ung,$^{7}$
    }
   \newauthor{%
    Prasenjit~Saha,$^{7}$
    Thomas~E.~Collett$^{8}$
    }
\medskip\\
$^1$\kipac\\
$^2$\oxford\\
$^3$\ipmu\\
$^4$\adler\\
$^5$\zooniverse\\
$^6$\oxfordeng\\
$^7$\zurich\\
$^8$\icg\\

}

\begin{document}

\date{to be submitted to MNRAS}
\pagerange{\pageref{firstpage}--\pageref{lastpage}}\pubyear{2014}

\maketitle

\label{firstpage}


\begin{abstract}

We describe \SW, a novel gravitational lens discovery service that yields
samples of high purity and completeness through crowd-sourced visual inspection.
Carefully produced colour composite images are displayed to volunteers via a
web-based classification interface, which records their estimates of the
positions of candidate lensed features. Images of simulated lenses, as
well as real
images which lack lenses, are inserted into the image stream at random
intervals; this training set is used to give the volunteers instantaneous
feedback on their performance, as well as to calibrate a model of the system
that provides dynamical updates to the probability that a classified image
contains a lens. Low probability systems are retired from the site periodically,
concentrating the sample towards a set of lens candidates. Having divided 160
square degrees of Canada-France-Hawaii Telescope Legacy Survey (\cfhtls) imaging
into some 430,000 overlapping 82 by 82 arcsecond tiles and displaying them on
the site, we were joined by around 37,000 volunteers who contributed 11~million
image classifications over the course of 8 months. This \StageOne search reduced
the sample to 3381 images containing candidates; these were then refined in
\StageTwo to yield a sample that we expect to be over 90\% complete and 30\%
pure, based on our analysis of the volunteers performance on training images.
We comment on the scalability of the \SW system to the wide field
survey era, based on our projection that searches of $10^5$ images could be performed
by a crowd of $10^5$ volunteers in 6 days.

\end{abstract}



\setcounter{footnote}{1}


\section{Introduction}
\label{sec:intro}


Strong gravitational lensing -- the formation of multiple, magnified images of
background objects due to the deflection of light by  massive foreground
objects -- is a very powerful astrophysical tool, enabling a wide range of
science projects. The image separations and distortions provide information
about the mass distribution in the lens \citep[e.g.][]{AugerEtal2010,
SonnenfeldEtal2012,MoreEtal2012,SonnenfeldEtal2015}, including on sub-galactic scales
\citep[e.g.][]{Dalal+Kochanek2002,VegettiEtal2010,HezavehEtal2013}. Any strong
lens can provide magnification of a factor of 10 or more, providing a deeper,
higher resolution view of the distant universe through these ``cosmic
telescopes'' \citep[e.g.][]{StarkEtal2008,NewtonEtal2011}. Lensed quasars
enable cosmography via the time delays between the lightcurves of multiple images \citep[e.g.][]{TewesEtal2013,SuyuEtal2013}, and study of the
accretion disk itself through the microlensing effect
\citep[e.g.][]{PoindexterEtal2008}. All of these investigations would
benefit from being able to draw from a larger and/or more diverse sample of lenses.

In the last decade the number of these rare cosmic alignments known
has increased by an order of magnitude, thanks to searches carried out in
wide field surveys, such as
CLASS \citep[e.g.]{BrowneEtal2003}, SDSS \citep[e.g.][]
{BoltonEtal2006,AugerEtal2010b,TreuEtal2011,InadaEtal2012,
HennawiEtal2008,BelokurovEtal2009,DiehlEtal2009,FurlanettoEtal2013},
\cfhtls \citep[e.g.][]{MoreEtal2012,GavazziEtal2014}, Herschel
\citep[][e.g.]{NegrelloEtal2014} and SPT \citep[e.g.][]{VieiraEtal2013}, among
others.  As the number of known lenses has increased, new types have been
discovered, leading to entirely new investigations. Compound lenses
\citep{GavazziEtal2008,CollettEtal2012} and lensed supernovae
\citep{QuimbyEtal2014,KellyEtal2015} are good examples of this.

Strong lenses are expensive to find, because they are rare. The highest purity
searches to date have made use of relatively clean signals such as the
presence of emission or absorption features at two distinct redshifts in the
same optical spectrum \citep[e.g.][]{BoltonEtal2004}, or the strong
``magnification bias'' towards detecting strongly-lensed sources in the sub-mm/mm
waveband \citep[e.g.][]{NegrelloEtal2010}. Such searches have to yield pure samples,
because they require expensive high resolution imaging follow-up; consequently
they have so far produced yields of only tens to hundreds of lenses. An alternative
approach is to search images already of sufficiently high resolution and colour
contrast, and confirm the systems as gravitational lenses by modeling the
survey data themselves \citep[][]{MarshallEtal2009}. Several square degrees of
HST images have been searched, yielding several tens of galaxy-scale lenses
\citep[e.g.][]{MoustakasEtal2007,Jackson2008,MoreEtal2011,
PawaseEtal2014}. Similarly, searches of over a hundred square degrees of CFHT
Legacy Survey (\cfhtls) ground-based imaging, also with sub-arcsecond image quality,
have revealed a smaller number of wider image separation group-scale systems
\citep[e.g.][]{CabanacEtal2007,MoreEtal2012}. Detecting galaxy-scale lenses
from the ground is difficult, but feasible albeit with lower efficiency and requiring
HST or spectroscopic follow-up to confirm the candidates as lenses
\citep[e.g.][]{GavazziEtal2014}.

How can we scale these lens searches up to imaging surveys that cover a hundred
times the sky area, such as the almost-all sky surveys planned with the Large
Synoptic Survey Telescope (LSST) and
Euclid?  There are two approaches to detecting lenses in imaging surveys. The
first one is robotic: automated analysis of the object catalogs and the survey
images. The candidate samples produced by these methods have, to date, only
reached purities of 1-10\%, with visual inspection
by teams of humans still required to reduce the robotically-generated
samples by factors of 10-100 \citep[see
e.g.][]{MarshallEtal2009,MoreEtal2012,GavazziEtal2014}.
In this approach, the image data may or may not be explicitly modelled
by the robots as if it contained a gravitational lens, but the visual inspection
can be thought of as a ``mental modeling'' step.  An inspector who classifies an
object as a lens candidate does so because the features in the image that they
see can be explained by a model, contained in their brain, of what gravitational
lenses do.  The second approach simply cuts out the robot middleman:
\citet{MoustakasEtal2007,FaureEtal2008,Jackson2008} and \citet{PawaseEtal2014}
all performed successful visual searches for lenses in HST imaging.

Until this problem is solved by machine learning tools,
at present visual image inspection seems unavoidable at some level when
searching for gravitational lenses. The technique has some drawbacks, however.
The first is that humans make mistakes. A solution to this is for the inspectors
to operate in teams, providing multiple classifications of the same images in
order to catch errors and correct them. Second, and relatedly, is that humans
get tired. With a well-designed classification interface, a human might be able
to inspect images at a rate of one astronomical object per second (provided the
majority are indeed uninteresting). At $10^4$ massive galaxies, and 10 lenses,
per square degree, visual lens searches in good quality imaging data are limited
to a few square degrees per inspector per day (and less, if more time is spent
assessing difficult systems). Scaling to thousands of square degrees therefore
means either robotically reducing the number of targets for inspection, or
increasing the number of inspectors, or both.

For example, a $10^4$ square degree survey containing $10^8$
photometrically-selected massive galaxies (and $10^5$ lenses) could only be
searched by 10 inspectors (at a mean rate of 1 galaxy per second and 3
inspections per galaxy) in about 5 years. Alternatively,  an automated system
could be asked to produce a much purer sample: if it was able to reach a purity
of 10\%, this would leave  $10^6$ lens candidates (100 targets per square
degree) to be visually inspected.  At this point the average visual
classification time per object could well be more like 10 seconds per object,
requiring the same  team of 10 inspectors to work full time for 20 weeks (to
provide 3 classifications  per lens between them). Neither of these may be the
most cost-effective or reliable strategy. Alternatively, a team of $10^6$
inspectors could, in principle, make the required $10^9$ image classifications,
$10^3$ each, in a few weeks; robotically reducing the target list would lead to
a proportional decrease in the required team size.

Systematic detection of rare astronomical objects by such ``crowd-sourced''
visual inspection has recently been demonstrated by the online citizen science
project \PH \citep{SchwambEtal2012}.
\PH was designed to enable the discovery of transiting exoplanets in data taken
by the Kepler satellite. A community of over 200,000 inspectors from the general
public found, after each undergoing a small amount of training, over 40 new
exoplanet candidates. They achieved this by visually inspecting the Kepler
lightcurves that were presented in a custom web-based classification interface
\citep{WangEtal2013}.
The older \GZ morphological classification
project \citep{LintottEtal2008} has also enabled the discovery of rare objects,
via its flexible inspection interface and discussion forum
\citep{LintottEtal2009}. Indeed, several of us (AV,CC,CM,EB,PM,LW) were
active in an informal \GZ gravitational lens search (Verma et al, in
preparation), an experience which led to the present hypothesis that a
systematic online visual lens search could be successful.

In this paper, we describe the \SW project, a web-based system conceived to
address the visual inspection problem in gravitational lens detection for future
large surveys by ``crowd-sourcing'' it to a community of citizen scientists.
Implemented as a Zooniverse \citep{SimpsonEtal2014}
project, it is designed to provide a {\it gravitational lens discovery service}
to survey teams looking for lenses in wide field imaging data. In a companion
paper \citep[][hereafter \PaperTwo]{MoreEtal2015} we will present the new
gravitational lenses discovered in our first lens search, and
begin to investigate the differences between lens detections made in \SW and
those made with automated techniques. Here though, we simply try to answer the
following questions:

\begin{itemize}

\item How reliably can we find gravitational lenses using the \SW
system? What is the completeness of the sample produced likely to be?

\item How noisy is the system? What is the purity of the sample
produced?

\item How quickly can lenses be detected, and non-lenses be rejected?
How many classifications, and so how many volunteers, are needed per target?

\item What can we learn about the scalability of the crowd-sourcing approach?

\end{itemize}

Our basic method in this paper is to analyze the performance of the \SW system
on the ``training set'' of simulated lenses and known non-lenses. This allows us
to estimate completeness and purity with respect to gravitational lenses that
have the same properties of the training set. In \PaperTwo, we carry out a
complementary analysis using a sample of ``known'' (reported in the literature)
lenses.

This paper is organised as follows.  In \Sref{sec:design} we introduce the \SW
classification interface and the volunteers who make up the \SW
collaboration,  explain how we use the training images, and describe our
two-stage candidate selection strategy. We then briefly introduce, in
\Sref{sec:data}, the particular dataset used in our first test
of the \SW system, and how we prepared the images prior to displaying them in
the web interface. In \Sref{sec:swap} we describe our methodology for
interpreting the classifications made by the volunteers, and then present the
results of system performance tests  made on the training images in
\Sref{sec:results}.  We discuss the implications of our results for future
lens searches in \Sref{sec:discuss} and draw conclusions in
\Sref{sec:conclude}.


\section{Experiment Design}
\label{sec:design}

The basic steps of a visual search for gravitational lenses are: 1) prepare
the images, 2) display them to an inspector, 3) record the inspector's
classification of each image (as, for example, containing a lens candidate or
not) and 4) analyze that classification along with all others to produce a
final candidate list. We describe step~1 in \Sref{sec:data}, and step~4 in
\Sref{sec:swap}. In this section we take a volunteer's eye view, and begin by
describing the \SW classification interface, the crowd of volunteers, and the
interactions between the two.

\begin{figure*}
\centering\includegraphics[width=0.9\linewidth]{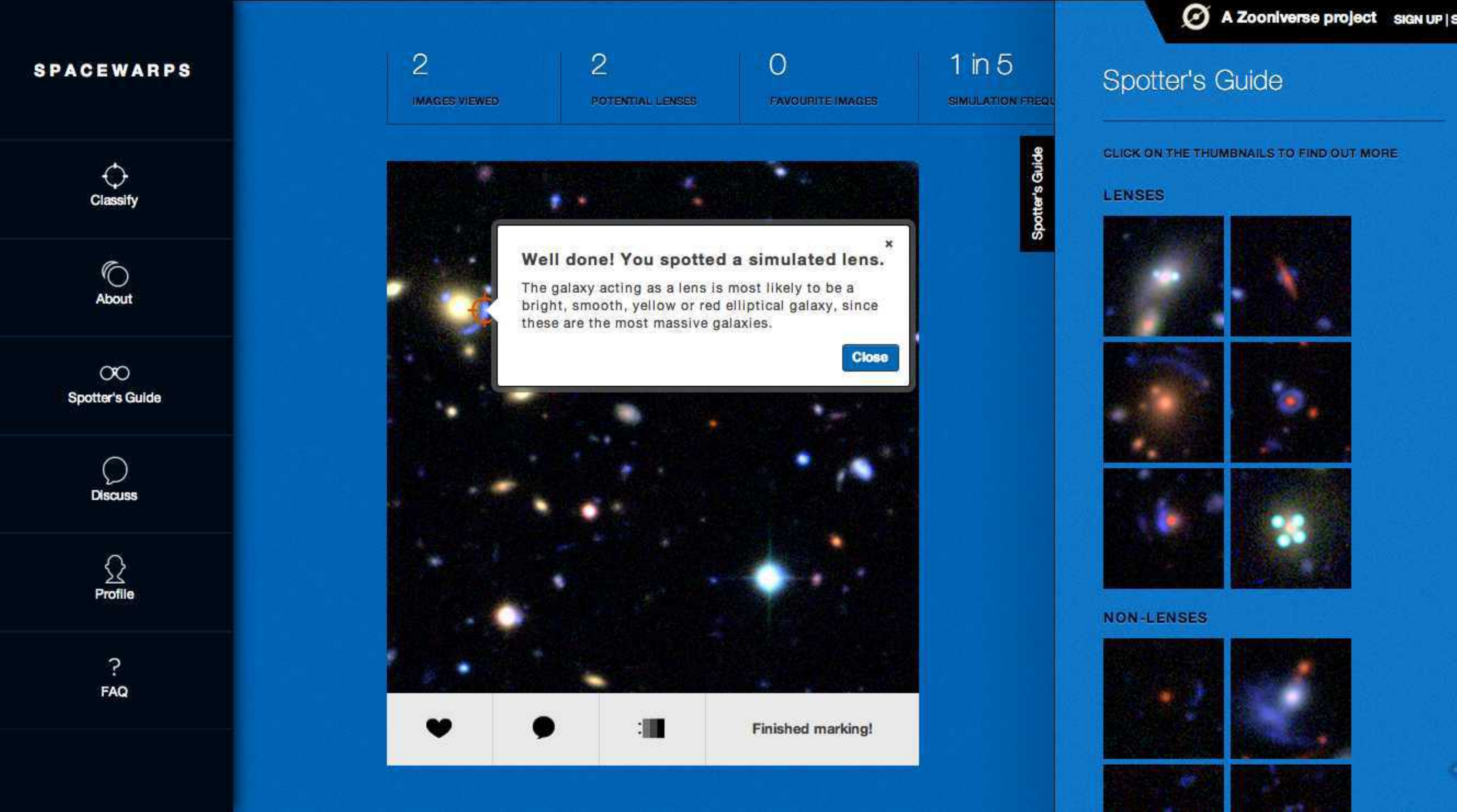}
\caption{Screenshot of the \SW classification interface at
  \texttt{http://spacewarps.org}.}
\label{fig:screenshot}
\end{figure*}


\subsection{The Classification Interface}
\label{sec:design:interface}

A screenshot of the \SW classification interface (CI) is shown in
\Fref{fig:screenshot}. The CI is the centrepiece of the \SW website,
\texttt{http://spacewarps.org}; the web application is written in coffeescript,
css and html and follows the general design of others written by the Zooniverse
team.\footnote{The \SW web application code is open source and can be accessed
from \texttt{https://github.com/zooniverse/Lens-Zoo}.  The project was renamed
during development to avoid the question of  who ``Len'' is.} The focus of the
CI is a large display of a pre-prepared PNG image of the ``subject'' being
inspected. When the image is clicked on by the volunteer, a marker symbol
appears at that position.  Multiple markers can be placed, and they can be moved or
removed by the classifier if they change their mind.

The next image moves rapidly in from a queue formed at the right hand side of
the screen when the ``Next'' or ``Finished Marking'' button is pressed (If a marker is placed somewhere in the image, the Next button changes to a ``Finished marking'' button). Gravitational lenses
are rare: typically, most of the images will not contain a lens candidate, and
these need to be quickly rejected by the inspector. The queue allows several
images to be pre-loaded while the volunteer is classifying the current subject,
and the rapid movement is deliberately designed to encourage volunteers to classify quickly. There is no ``back'' button in the CI: each volunteer may only classify a given
subject once and subjects cannot be returned to after the `Next or Finished marking' button has been pressed (this can be a source of missed lens candidates when classifying at speed). Note that by `classify' we mean that we interpret no markers being placed as a rejection, and placing at least one marker to mean a possible lensing event has been identified.
After each classification, the positions of any markers are written out to the
classification database, in an entry that also stores metadata including the ID
of the subject, the username of the volunteer (or IP address if thy are not
logged in), and a timestamp.

For the more interesting subjects, the CI offers two features that enable their
further investigation. The first is the ``Quick Dashboard''
(QD), a more advanced image viewer. This allows the viewer to compare three
different contrast and colour balance settings, to help bring out subtle
features, and to pan and zoom in on interesting regions of the image to assess
small features. Markers can be placed in the Quick Dashboard just the same as
in the main CI image viewer. The second is a link to that subject's page in
the project discussion forum (known as
``\Talk\footnote{\texttt{http://talk.spacewarps.org}}'').
Here, volunteers can discuss the features they have seen either
before they submit their classification, or after, if they ``favourite'' the
subject. The option of reading other opinions on any given image or lens candidate before submitting a classification means that
the classifications may not be strictly independent; however, the advantage of this
system is that volunteers can learn from others what constitutes a good lens
candidate (we are not able to track who visited talk before pressing ``Finished marking''). This is not, however, the primary educational resource; we describe
the explicit training that we provide for the volunteers in the next section.
The FITS images of the subject can be further explored with Zooniverse
``Dashboard Tools,'' which include a
more powerful image viewer that enables dynamic
variation of the colour balance and contrast (stretch), and for any given view
to be shared via unique URL back to \Talk. The FITS image viewing in both the
Quick Dashboard and the Dashboard Tools are enabled by the javascript library
\fitsjs \citep{fitsjs}.


\subsection{Training}
\label{sec:design:training}

Gravitational lenses are typically unfamiliar objects to the general public.
New volunteers need to learn what lenses look like as quickly as possible, so
that they can contribute informative classifications. They also need to learn
what lenses do not look like, in order to reduce the false positive detection
rate. There are three primary mechanisms in the \sw system for teaching the
volunteers what to look for. These are, in the order in which they are
encountered, an inline tutorial, instant feedback on ``training images''
inserted into the stream of images presented to them, and a ``Spotter's
Guide.'' As well as this, we provide ``About'', ``FAQ'' and ``Science'' pages
explaining the physics of gravitational lensing. While we expect that the
insight from this static material should help volunteers make sense of the
features in the images, we focus on the more dynamic, activity-based training
early, when engaging new volunteers to participate takes priority.

\subsubsection{Inline Tutorial}

New volunteers are welcomed to the site with a very short tutorial, in which
the task is introduced, a typical image containing a simulated lens is
displayed, and the marking procedure walked through, using pop-up message
boxes. Subsequent images gradually introduce the more advanced features of the
classification interface (the QD and \Talk buttons), also using pop-up
messages. The tutorial was purposely kept as short as possible so as to
provide the minimal barrier to entry.

\subsubsection{Training Subjects and Instant Feedback}

The second image viewed after the initial tutorial image is already a survey
image, in order to get the volunteers engaged in the real task as quickly as
possible. However, training then continues, beyond the first image tutorial,
through ``training subjects'' inserted randomly into the stream.  These
training subjects are either simulated lenses (known as ``sims''), or survey
images that were expert-classified and found not to contain any lens
candidates (these images are known as ``duds''). The tutorial explains that
the volunteers  will be shown such training images. They are also informed
that they will receive instant feedback about their performance after
classifying (blindly) any of these training subjects. Indeed, after a volunteer
finishes marking a training subject, a pop-up
message is generated, containing either congratulations for a successful
classification (for example, ``Well done! You spotted a simulated lens,'' as
in \Fref{fig:screenshot}) or feedback for an unsuccessful one (for
example, ``There is no gravitational lens in this field!''). A successful
classification of a simulated lens requires a marker to be placed fairly
precisely on the lensed features, so as to avoid misinforming the classifier.
The pixels in the lensed features are flagged via a mask contained in the
PNG image transparency ``alpha'' channel. (The mask is not visible in the image.)

\sw can be viewed as a {\it supervised learning} system,
because we include a training set of images in amongst the survey ``test''
images. Consequently,
we should expect to find lens candidates that look like the sims that we put in
the training set, and so the design of this training set is quite important. The
training images must look realistic, in two ways. First, apart from the
simulated lenses they contain, they must look like all the other survey images --
otherwise they could be identified as training images without the volunteer
learning anything about what lenses look like. For this reason, we make the sims
by {\it adding in} simulated lensed features to real images drawn from the
survey, and we make the duds by simply inspecting and choosing from a small
sample of randomly-selected survey images. It also means that each dataset
presented for inspection in the \sw interface needs to come with its own
training set.

Second, the lensed features themselves must be realistic; if they did not
resemble real lenses, we could not expect to find real lenses in the test
images. The details of how we generated the sims for the \cfhtls project are
given in \PaperTwo, where we also carry out performance tests relative to real
lens candidates in the survey fields. Here, we merely note the following
aspects to place this paper's results in context.

We began by selecting massive galaxies and groups in the \cfhtls catalogs,
by colour, brightness and photometric redshift. We then assigned plausible mass
distributions to these ``potential lenses:'' a singular isothermal ellipsoid
model for each massive galaxy, and an additional NFW profile dark matter halo
for groups and clusters. The mass distributions were centred, elongated and
aligned with the measured optical properties of the galaxies present. We then
drew a source from a plausible background population, of either galaxies (for
the group or galaxy lenses) or quasars (for the galaxy-scale lenses). The
source redshift, luminosity and size distributions were all chosen to match
those observed, while the source abundance was artificially increased such that
each potential lens had a high probability of having a source occur at a
position that will lead to it being highly magnified. Source galaxies were
given plausible surface brightness profiles and ellipticities. We then computed
the predicted multiple images of the source using the \gravlens ray-tracing
code \citep{Keeton2000}.

At this point, we applied several cuts to the simulated lens pool, in order to
reject sims that were likely to be difficult to spot. While this ensured
that the sims had high educational value, it means that we
should not necessarily expect to be able to
detect real lens candidates that would not pass these cuts.
However, human classifiers do possess the key advantage of being
able to {\it imagine} lens systems beyond what they were shown during their training.
Indeed, any real lens candidates we find with
properties outside the ranges of the training set could be of particular
interest, if the training set is indeed completely representative of lenses that
we already know about.
\citet{GeachEtal2015}, for example, report the discovery of a
red lensed arc  by volunteers who had only been trained primarily on blue/green
examples. We investigate this aspect of the \SW approach further in \PaperTwo.
In the present project, the most important selection cuts we made were to only
keep sims with a) second brightest image having $i < 23$ and b) total
combined image magnitude fainter than 19--20 (\PaperTwo, Table~1). The
combination of these two reductions resulted in a sample of visible lensed image
configurations.

Volunteers were initially shown training images at
a mean frequency of two in five. Subjects were drawn at random from a pool
consisting of (at first) 20\% sims, 20\% duds, and 60\% test images,
and such that no volunteer ever
saw a given image more than once. As the number of classifications made by the
volunteer increased, the training frequency was decreased from 40\% to
$2/(5\times2^{(\textrm{int}(N_c/20)+1)/2})$, $\approx 30\%$ for the second 20
subjects, $20\%$ for the third 20 subjects, and the minimum rate of 10\%
after that. We did not reduce the training frequency to zero because we wanted
to ensure that the inspectors remained alert.

This training regime meant that in the first 60 images viewed, each volunteer
was shown (on average) 9 simulated gravitational lenses (as well as 9
pre-selected empty ``dud'' fields). This is a much higher rate than the natural
one: to try and avoid this leading to over-optimism among the inspectors (and a
resulting high false positive rate), we displayed the current ``Simulation
Frequency'' on the classification interface (``1 in 5'' in
\Fref{fig:screenshot}) and maintained the consistent theme in the feedback
messages that lenses are rare.

In Figures~\ref{fig:training-gallery:sims}~and~\ref{fig:training-gallery:duds}
we show example training images from this first \SW project (\Sref{sec:data}
below).

\subsubsection{Spotter's Guide}
\label{sec:design:training:guide}

The instant feedback provides real-time educational responses to the volunteers
as they start classifying; as well as this dynamic system, \SW provides a static
reference work for volunteers to consult when in doubt about how to perform the
task. This ``Spotter's Guide'' is a set of webpages showing example lenses, both
real and simulated, and also some common false positives, drawn from the pool of
survey images. The false positives were identified during the selection of the
``dud'' training images (previous section). For easy reference, the lenses are
divided by type (for example, ``Lensed Galaxies,'' ``Lensed Quasars'' and
``Cluster Lenses''), as are the false positives (for example, ``Rings and
Spirals,'' ``Mergers,'' ``Artifacts'' and so on). The example images are
accompanied by explanatory text. The Spotter's Guide is reached via a button on
the left hand side, or the hyperlinked thumbnail images of the ``Quick
Reference'' provided on the right hand side, of the classification interface.

Most of the text of the Spotter's Guide focuses on the kinds of features that
gravitational lenses do or don't produce. To complement this,
the website ``Science'' section contains a very brief introduction to how
gravitational lensing works, which is fleshed out a little on the ``FAQ'' page.
The FAQ also contains answers to questions about the interface
and the task set.


\subsection{Staged Classification}
\label{sec:design:stages}

We now describe briefly the two-stage strategy that we employed in this first
project: initial classification (involving the rejection of very large numbers
of non-lenses), and refinement (to further narrow down the sample). The web
application was reconfigured between the two stages, to assist in their
functioning.

\subsubsection{\StageOne: Initial Classification}

The goal of the \StageOne classification was to achieve a high rejection rate,
while maintaining high completeness.  In this mode, therefore, the pre-loading
of images was used to make the sliding in of new subjects happen quickly, to
provide a sense of urgency: initial classification must be done fairly
quickly for the search to be completed within a reasonable time period.  We
expect some trade-off between speed and accuracy: we return to this topic in the
results section below. Completion of the search requires subjects to be
``retired'' over time, as a result of their being classified. We did this by
analyzing the classifications on a daily basis, as described in
\Sref{sec:swap} below. As subjects were retired, new ones were ingested into
the web app for classification. This means that the discovery of lens
candidates in \StageOne is truly a community effort: to detect a lens candidate,
many non-lenses must first be rejected, and several classifications by
different inspectors are needed in either case.

The \StageOne training set was chosen to be quite clear cut, in order to err on
the side of high completeness. When defining the
training duds, we discarded anything that could be considered a lens candidate
(see \Sref{sec:design:training:guide}). This meant that objects that look
similar to lenses, such as galaxy mergers, tidal tails and spiral arms, pairs of
blue stars and so on, were specifically excluded from the training set, and
therefore we expect some of those types of object to appear in the \StageOne
candidate list. As described above, the training sims for \StageOne were also
selected to be relatively straightforward to spot.

\subsubsection{\StageTwo: Refinement}
\label{sec:design:stages:two}

\begin{figure*}
\begin{minipage}[b]{0.24\linewidth}
\centering\includegraphics[width=\linewidth]{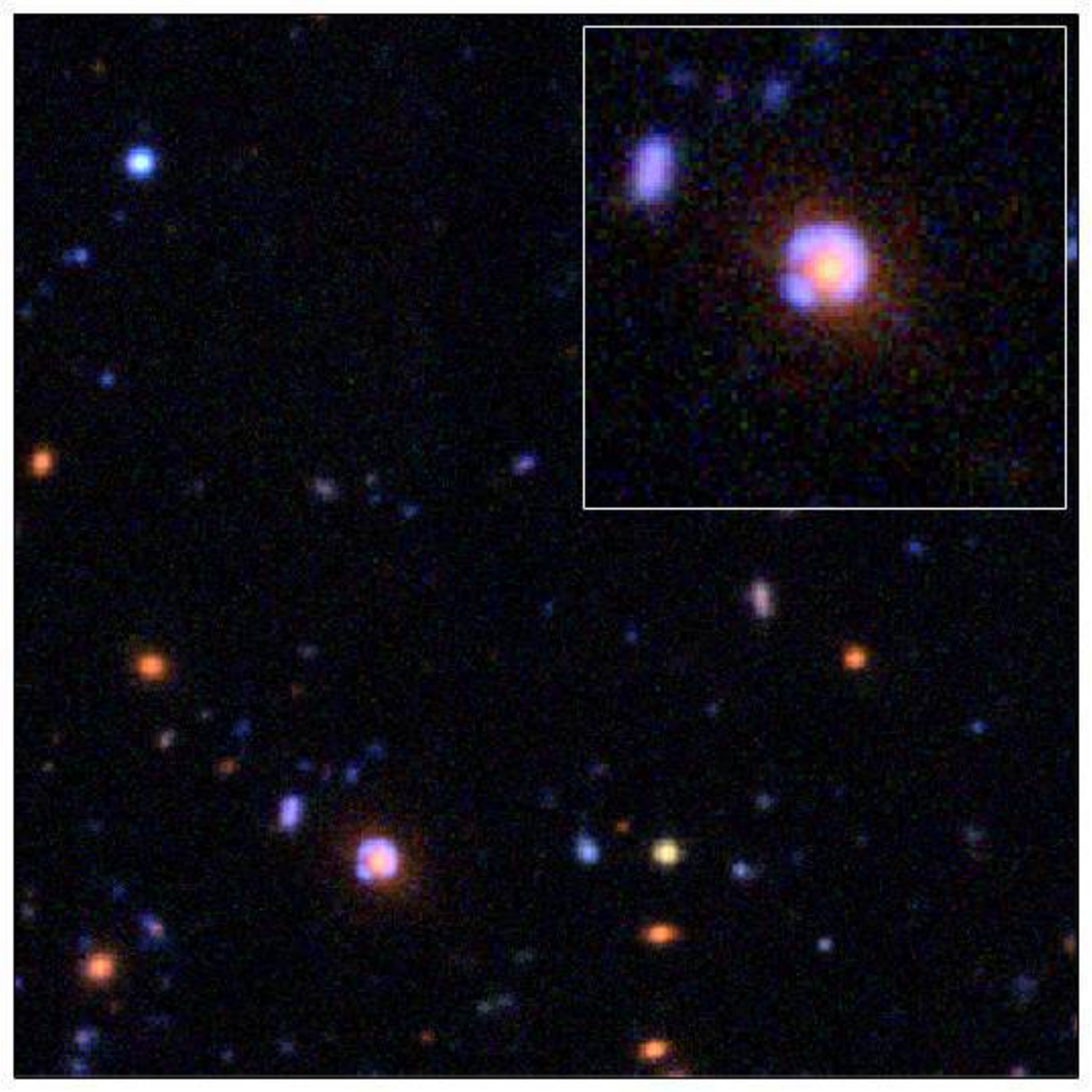}
\end{minipage} \hfill
\begin{minipage}[b]{0.24\linewidth}
\centering\includegraphics[width=\linewidth]{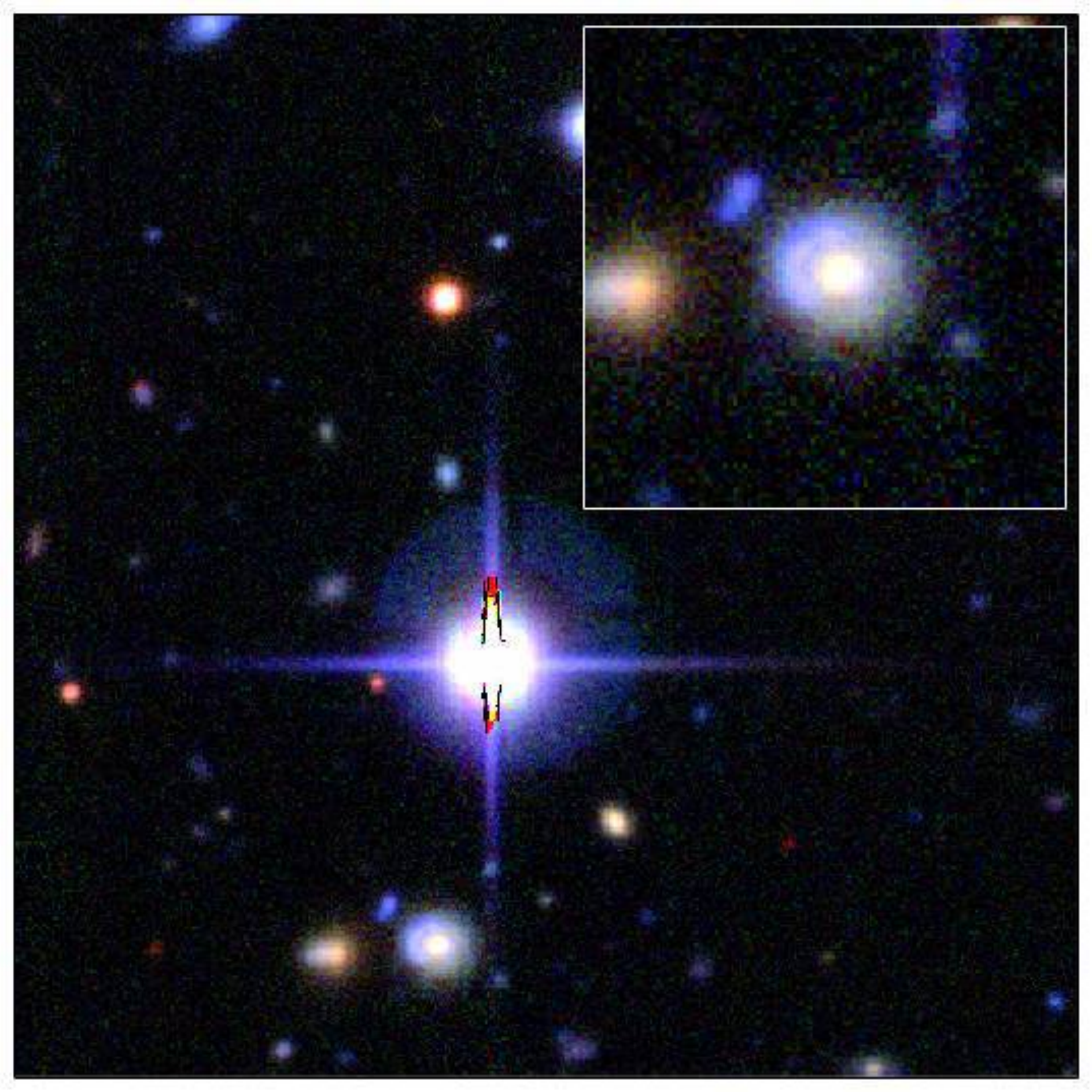}
\end{minipage} \hfill
\begin{minipage}[b]{0.24\linewidth}
\centering\includegraphics[width=\linewidth]{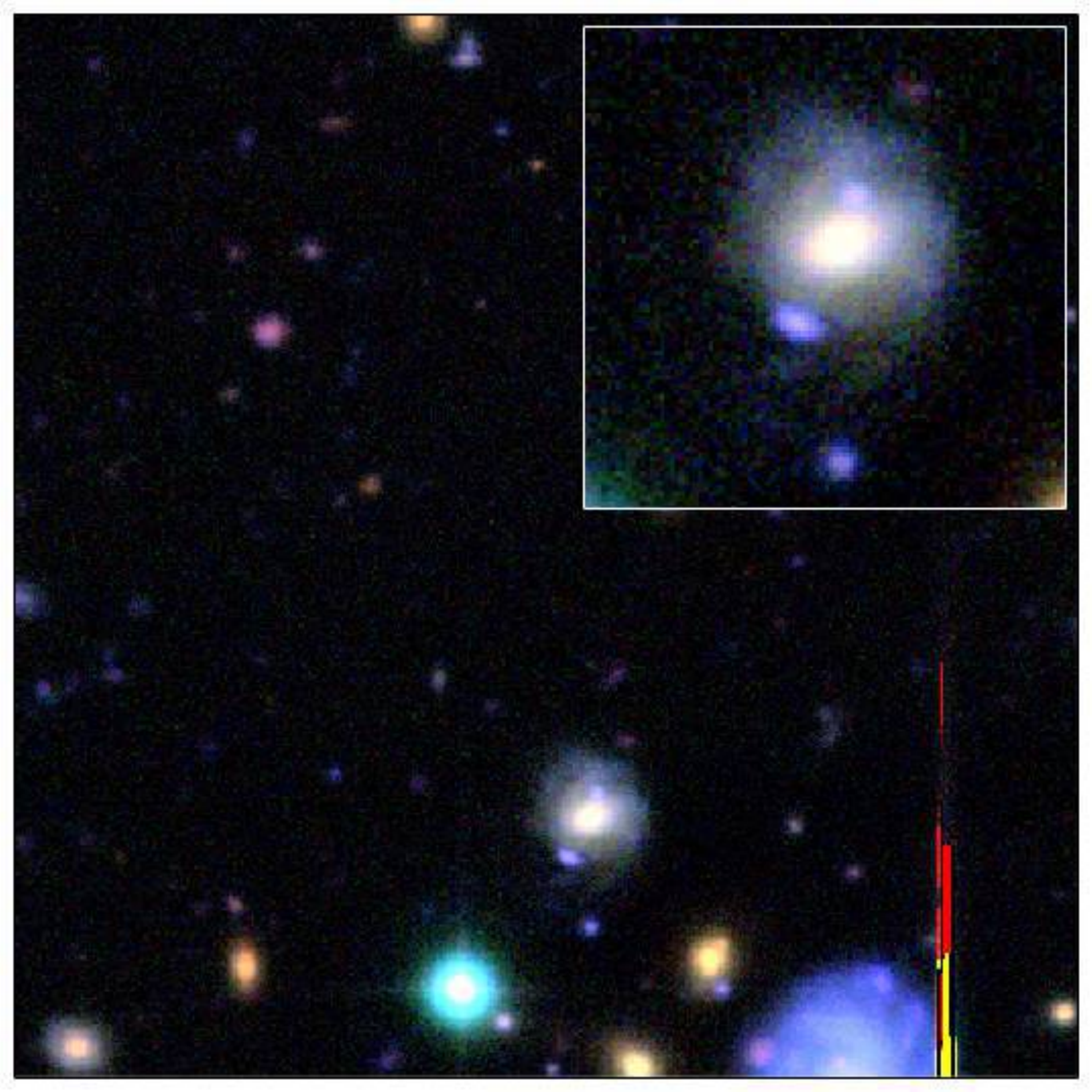}
\end{minipage} \hfill
\begin{minipage}[b]{0.24\linewidth}
\centering\includegraphics[width=\linewidth]{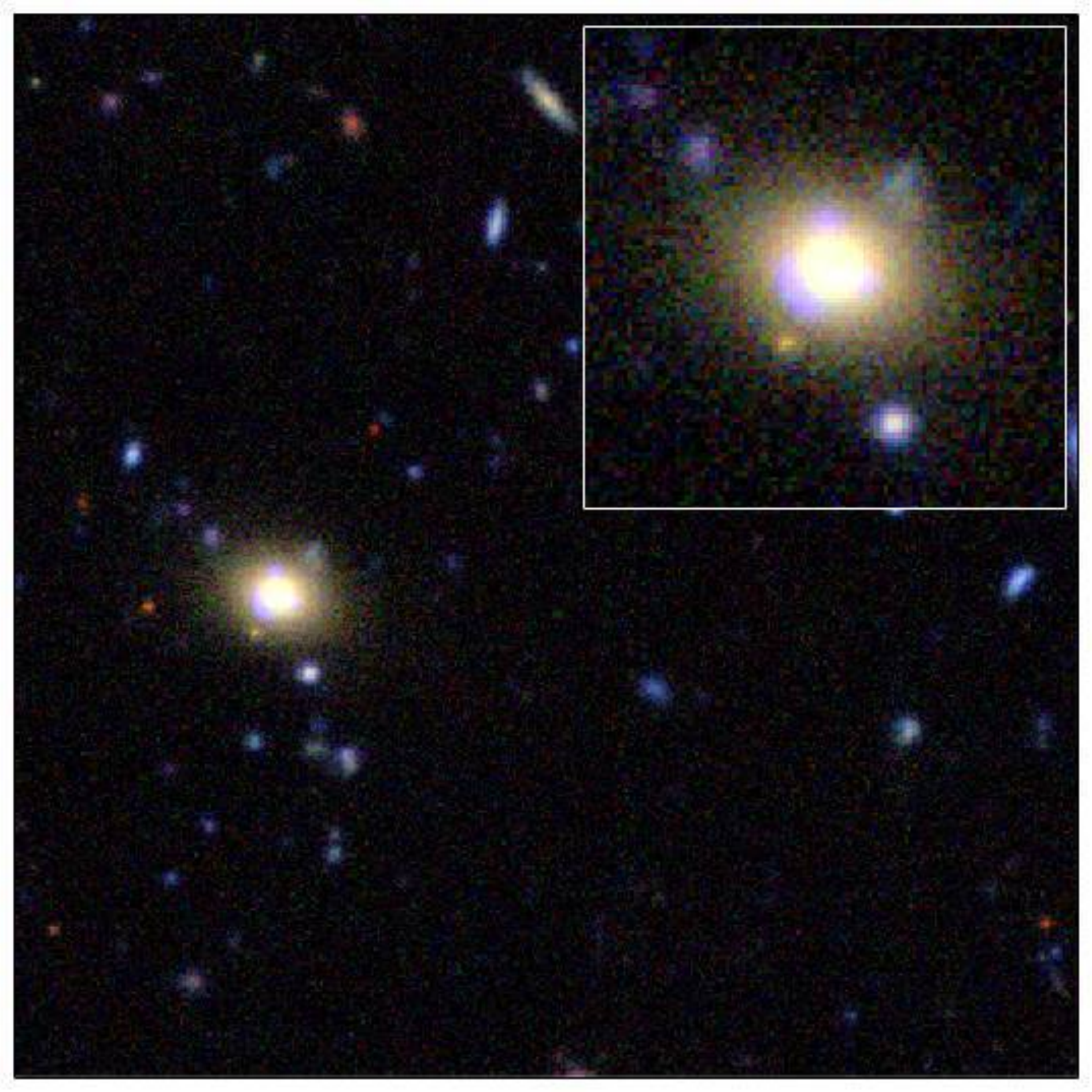}
\end{minipage} \hfill
\caption{Typical \SW ``sims'' from the \StageTwo training set.
The top right-hand corner insets indicate the
location of the simulated lens in each of these training images. Volunteers needed to
click on these specific features in order to make a correct classification.}
\label{fig:training-gallery:sims}
\end{figure*}

\begin{figure*}
\begin{minipage}[b]{0.24\linewidth}
\centering\includegraphics[width=\linewidth]{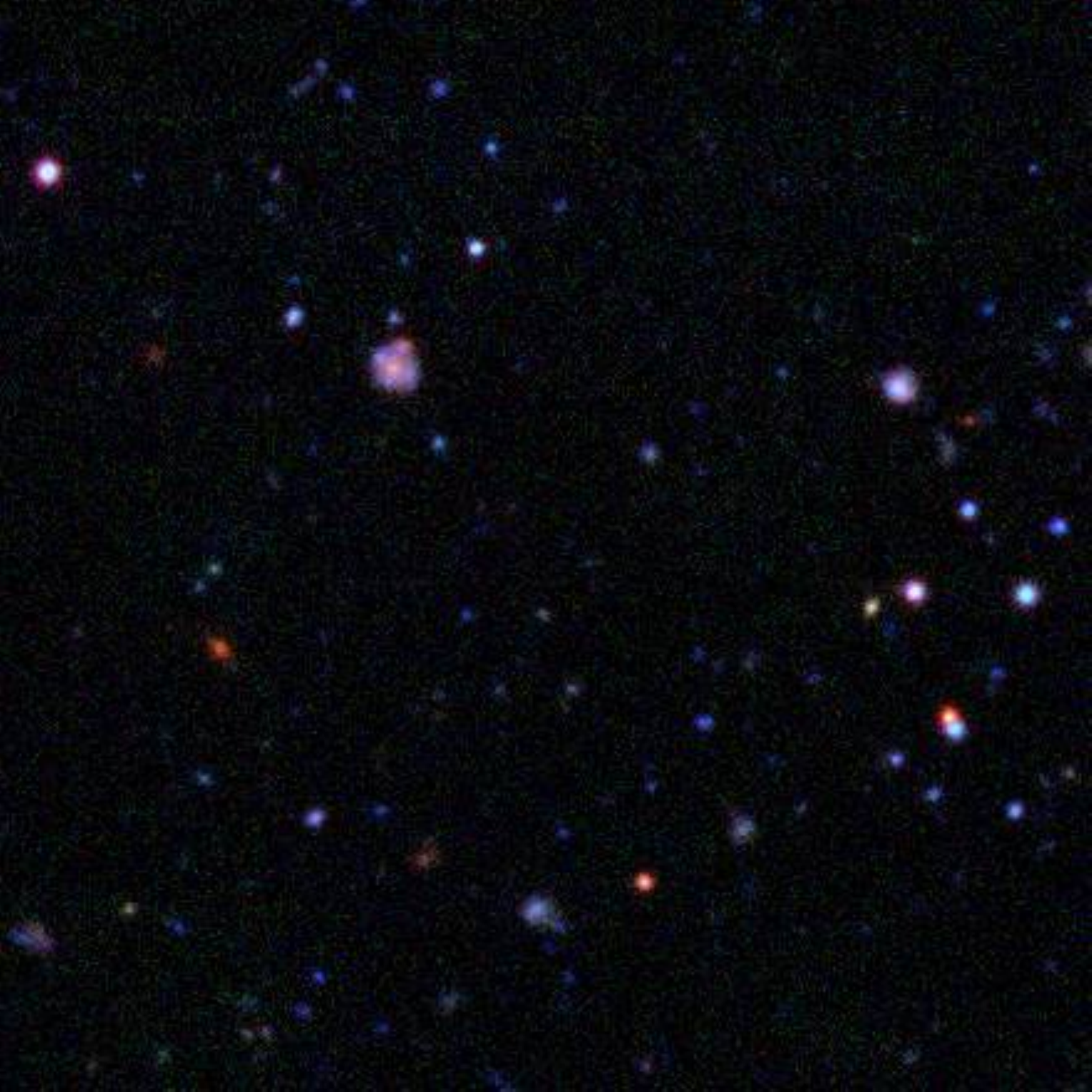}
\end{minipage} \hfill
\begin{minipage}[b]{0.24\linewidth}
\centering\includegraphics[width=\linewidth]{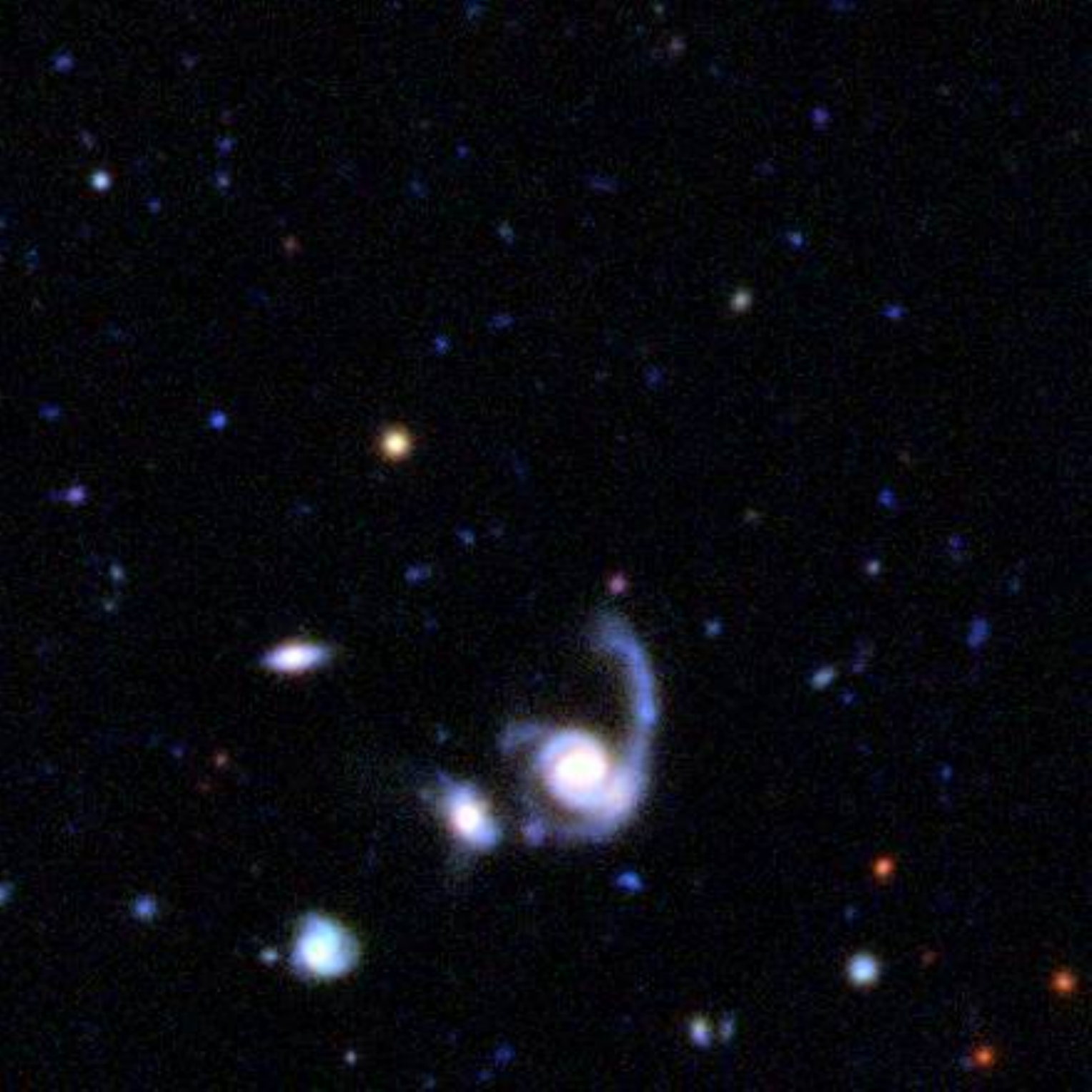}
\end{minipage} \hfill
\begin{minipage}[b]{0.24\linewidth}
\centering\includegraphics[width=\linewidth]{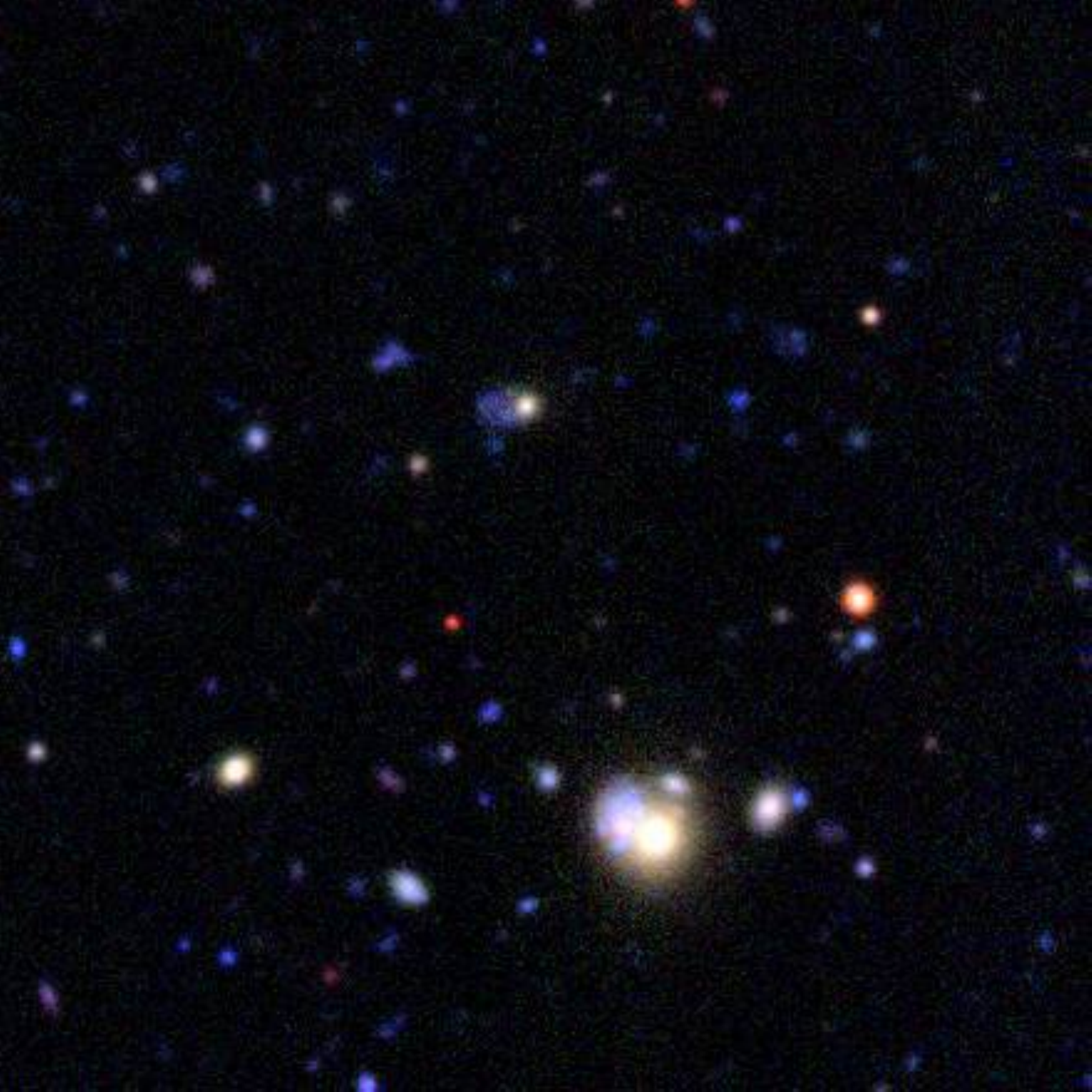}
\end{minipage} \hfill
\begin{minipage}[b]{0.24\linewidth}
\centering\includegraphics[width=\linewidth]{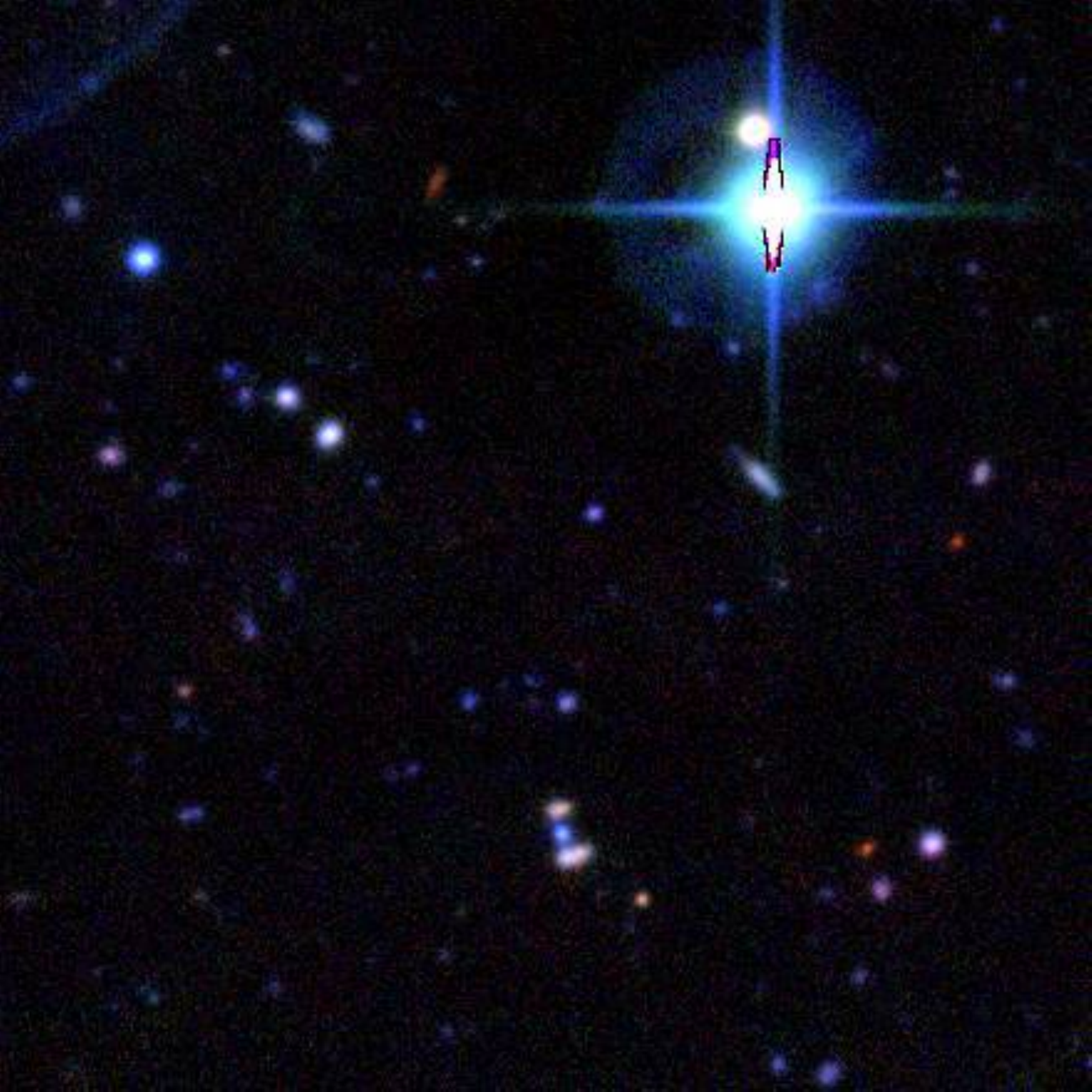}
\end{minipage} \hfill
\caption{Typical \SW ``duds'' from the \StageTwo training set. All of these
subjects were correctly classified by the community as not likely to contain
gravitational lenses.} \label{fig:training-gallery:duds}
\end{figure*}

The design of a \StageOne classification task, and its training set, should
lead to a sample of lens candidates that has high completeness but may have low
purity. To refine this sample to higher purity, we need to reject more
non-lenses, which means providing the volunteers with a more realistic and
challenging training set as they re-classify it. The more demanding \StageTwo
training set was generated as follows. The \StageTwo duds were selected
from a small random subset of the \StageOne candidates (i.e., the \StageTwo duds
were expert-defined \StageOne false positive detections), while the \StageTwo
sims were chosen to be a subset of the \StageOne sims, none of which were deemed
``obvious'' by the same expert classifiers. This meant that the \StageTwo sims
had fainter and less well-separated image features than in the \StageOne
training set.  Figures~\ref{fig:training-gallery:sims}
and~\ref{fig:training-gallery:duds} show some example images from the resulting
\StageTwo training set.

We also attempted to encourage discernment by changing the look and feel of the
app, slowing down the arrival of new images, and switching the background colour
to bright orange to make it clear that a different task was being set. The
frequency with which training images were shown was fixed at 1 in 3. Finally,
the Spotter's Guide was upgraded to include more examples of various possible
false positives, divided into sub-classes. We did not retire any subjects during
\StageTwo classification, instead continuing to accumulate classifications
through till the end of a fixed 4-week time period.


\section{Data}
\label{sec:data}

We refer the reader to \PaperTwo for the details of the particular set of
imaging survey data used in this first \SW project. Here, we summarize very
briefly the choices that were made, in order to provide the context for our
general description and illustrations of the \SW system.


\subsection{The CFHT Legacy Survey}
\label{sec:data:CFHTLS}

The four CFHT Legacy Survey\footnote{\texttt{http://www.cfht.hawaii.edu/Science/CFHTLS/}}
\citep[\cfhtls,][]{Gwyn2012} ``Wide'' fields cover a total of approximately
160 square degrees of sky (after taking into account tile overlaps).
With high and homogeneous image quality (the mean seeing in the $g$-band is
0.78''), and reaching limiting magnitudes of around 25 across the $ugriz$ filter
set, this survey has yielded several dozen new gravitational lenses on both
galaxy and group scales
\citep{GavazziEtal2014,SonnenfeldEtal2013a,CabanacEtal2007,MoreEtal2012}. The
quality of the data, combined with the presence of these comparison ``known
lens'' samples, makes this a natural choice against which to develop and test
the \SW system. The \cfhtls is also well-representative of the data quality
expected from several next-generation sky surveys, such as DES, KiDS, HSC and
LSST. We use the stacked images
from the final T0007 release taken from the Terapix
website\footnote{\texttt{ftp://t07.terapix.fr/pub/T07/}} for this
work.

In order to investigate the completeness of the previous semi-automated lens
searches in the \cfhtls area, we designed a ``blind search'' as follows.
We divided the
\cfhtls pointings into some 430,000 equal size, overlapping tiles, approximately
82~arcsec on a side. We refer to these images as ``test images.'' The ``training
images'' were derived from a small subset of these, as  discussed above. In
future, larger area, projects we expect to implement a somewhat different
strategy of producing image tiles centred on particular pre-selected
``targets,'' which might make for a more efficient (if less complete) survey.
However, we do not expect the performance of citizen image inspectors to change
significantly between these strategies: to first order, both strategies require
the inspectors to learn what lenses look like, and then search the presented
images for similar features.


\subsection{Image Presentation}
\label{sec:data:display}

The \cfhtls $g$, $r$ and $i$-band images have the greatest average
depth and highest average image
quality of the survey,
and we chose to focus on this subset. (The $u$ and $z$-band images
were also made available for perusal in \Talk). We made colour composite PNG
format images following the prescription of \citet{LuptonEtal2004} \citep[with
extensions by][and some particular choices of our own]{WherryEtal2004}, using
the \humvi software.\footnote{The \humvi colour image composition code used in
this work is open source and available from
\texttt{http://github.com/drphilmarshall/HumVI}} Specifically, we first rescaled
the pixel values of each image, in the notation of \citeauthor{LuptonEtal2004},
into flux units (``picomaggies''), via the image AB zeropoint $m_0$ and the
pixel value calibration factor $f$  given by $\log_{10}{f} = 0.4(30.0 - m_0)$.
We then multiplied these calibrated images by further aesthetic ``scales''
$s_{i,r,g}$ before computing the total intensity image $I$ and applying an
arcsinh stretch. (The scales are re-normalized to sum to one on input.) Thus,
the red ($R$), green ($G$) and blue ($B$) channel images correspond to the
\cfhtls $i$, $r$ and $g$-band images in the following way:
\bea
I & = & (i \cdot s_i + r \cdot s_r + g \cdot s_g), \notag \\
R & = & i \cdot s_i \cdot \frac{{\rm asinh}\left(\alpha \cdot Q \cdot I\right)}{Q \cdot I}, \notag \\
G & = & r \cdot s_r \cdot \frac{{\rm asinh}\left(\alpha \cdot Q \cdot I\right)}{Q \cdot I}, \notag \\
B & = & g \cdot s_g \cdot \frac{{\rm asinh}\left(\alpha \cdot Q \cdot I\right)}{Q \cdot I}. \notag \\
\eea
We chose to allow the composite image formed from these channel images to
saturate to white: any pixels in any of the channel images lying outside the
range 0 to 1 was assigned the value 0 or 1 appropriately. This was not the
recommendation of \citet{LuptonEtal2004}, but we found it to still give very
informative but also familiar-looking astronomical images.

The non-linearity parameters $Q$ and $\alpha$ control the brightness and
contrast of the images. We first tuned $\alpha$ (which acts as an additional
scale factor) until the background noise was just visible. Then we tuned the
colour scales $s$ to find a balance between exposing the low surface
brightness blue features (important for lens spotting!), and having the noise
appear to have equal red, green and blue components. Non-linearity sets in at
about $1/(Q \alpha)$ in the  scaled intensity image. Finally, tuning $Q$ at
fixed $\alpha$ determines the appearance of bright galaxies, which we need to
be suppressed enough to allow the low surface brightness features to show
through, but not so much that they no longer looked like massive galaxies.

These parameters were then fixed during the production of all the tiles, in
order to allow straightforward comparison between one image and another, and
for intuition to be built up about the appearance of stars and galaxies across
the survey. (Alternative algorithms, such as adjusting the stretch and scale
dynamically according to, for example, the root-mean-square pixel value in
each image, can lead to better presentation of bright objects, but in doing so
they tend to hide the faint features in those images: we needed to optimize
the detectability of these faint features.)   Examples of \cfhtls training set
images prepared in this way can be seen in Figures~\ref{fig:training-gallery:sims}
and~\ref{fig:training-gallery:duds}. We also defined two alternative sets of
visualization parameters, to display a ``bluer'' image and a ``brighter''
image in the classification interface Quick Dashboard
(\Sref{sec:design:interface}). The QD code performs the same image composition
as just described, but dynamically on FITS images in the browser. These FITS
images are also available for viewing in \Talk, via the main Zooniverse
Dashboard image display tool, which again offers the same stretch settings, as
a starting point for image exploration. Our parameter choices are given in
\Tref{tab:data:humvi}.

\begin{table}
\begin{center}
\caption{Example \humvi image display parameters, for the \cfhtls images.}
\label{tab:data:humvi}
\begin{tabular}{cccc}
  \hline
  \hline {Image} & Scales ${s_i,s_r,s_g}$ & $\alpha$ & $Q$ \\
  \hline
    ``Standard'' &    ${0.4,0.6,1.7}$     &   0.09   & 1.0 \\
    ``Brighter'' &    ${0.4,0.6,1.7}$     &   0.17   & 1.0 \\
       ``Bluer'' &    ${0.4,0.6,2.5}$     &   0.11   & 2.0 \\
  \hline \hline
\end{tabular}
\medskip\\
\end{center}
\end{table}


\section{Classification Analysis}
\label{sec:swap}

Having described the classification interface, the training images and the
test images,  we now outline our methodology for interpreting the
classifications made by the
volunteers, and describe how we
applied this methodology in the two classification stages of the \cfhtls
project in the \SW Analysis Pipeline (SWAP) code.\footnote{The open source
SWAP code is available from
\texttt{https://github.com/drphilmarshall/SpaceWarps}}

Each classification made is logged in a database that stores the subject ID,
volunteer ID, a timestamp and the results of each classification: the image
pixel-coordinate positions of every marker placed.  The ``category'' of subject --
whether it is a ``training'' subject (a simulated lens or a known non-lens) or a
``test'' subject (an unseen image drawn from the survey) -- is also recorded.
For training subjects, we also store the ``kind'' of the subject as
a lens (``sim''), or a non-lens (``dud''), and also the ``flavour'' of lens
object if one is present in the image (``lensed galaxy'', ``lensed quasar'' or
``cluster lens'').  This information is used to provide the instant feedback, but
is also the basic data used in a probabilistic classification of every
subject based on all image views to date. Not all
volunteers register with the Zooniverse (although all are prompted to do so);
in these cases we record their IP addresses as substitute IDs.

While the \SW web app was live, and classifications were being made, we
performed a daily online analysis of the classifications,
updating a probabilistic model of every
anonymous volunteer's data, and also updating the posterior probability that
each subject (in both the training and test sets) contains a lens.
This gave us a dynamic estimate of the posterior probability for  any
given  subject being a lens, given all classifications of it to date. Assigning
thresholds in this lens probability then allowed us to make good decisions about
whether or not to retire a subject from the system, in order to focus attention
on new images.

The details of how the lens probabilities are calculated are given below.
In summary:
\begin{itemize}

\item Each volunteer is assigned a simple software agent, characterised by a
confusion matrix~$\CM$. The two independent elements of this matrix are the
probabilities, as estimated by the agent, that the volunteer is going to be 1)
correct when they report that an image contains a lens when it really does
contain a lens, $\PL = \pr(\saidLENS|\LENS,T)$, and 2) correct when they report that
an image does not contain a lens when it really doesn't contain a lens,
$\PD = \pr(\saidNOT|\NOT,T)$. The confusion matrix contains all the information
the agent has about how good its volunteer is at classifying images.

\item Each agent updates its confusion matrix elements based on the number of
times its volunteer has been right in each way (about both the sims and the
duds) while classifying subjects from the training set, accounting for noise
early on due to small number statistics: $T$~is the set of all training images
seen to date.

\item Each agent uses its confusion matrices to update, via Bayes' theorem,
the probability of an image from the test set containing a lens,
$\pr(\LENS|C,T)$, when that image is classified by its volunteer. ($C$ is the
set of all classifications made of this subject.)

\end{itemize}

For a detailed derivation of this analysis pipeline, please continue reading
through \Sref{sec:swap:details} below. Alternatively,
\Sref{sec:swap:practicalities} contains illustrations of the calculation.


\subsection{SWAP: the \SW Analysis Pipeline}
\label{sec:swap:details}

Our aim is to enable the construction of a sample of good lens candidates.
Since we aspire to making logical  decisions, we define a  ``good candidate''
as one which has a high posterior probability of being a lens, given the data:
$\pr(\LENS|\data)$. Our problem is to approximate this probability. The data~$\data$
in our case are the pixel values of a colour image. However, we can greatly
compress these complex, noisy sets of data by asking each volunteer what they
think about them. A complete  classification in \SW consists of a set of
Marker positions, or none at all. The null set encodes the statement from
the volunteer that the image in question is $\saidNOT$ a lens, while the
placement of any  Markers indicates that the volunteer considers this image to
contain a $\saidLENS$.  We simplify the problem by only using the Marker
positions to assess whether the volunteer  correctly assigned the
classification $\saidLENS$ or $\saidNOT$ after viewing (blindly) a member of
the training set of subjects.

How should we model these compressed data? The circumstances of each
classification are quite complex, as are the human classifiers themselves: the
volunteers learn more about the problem as they go, but also inevitably make
occasional mistakes (especially when classifying at high speed).
To cope with this uncertainty, we assign a
simple software {\it agent} to partner each volunteer. The agent's task is to
interpret their volunteer's classification data as best it can, using a model
that makes a number of necessary approximations. These interpretations will
then include uncertainty arising as a result of the volunteer's efforts and
also the agent's approximations.
The agent will be able to predict, using its model, the probability of a test
subject being a $\LENS$ or an empty field given both its volunteer's classification and its
volunteer's past experience. In this section we describe how these agents work,
and other aspects of the \SW analysis pipeline (SWAP).

\subsection{Agents and their Confusion Matrices}
\label{sec:swap:details:probabilities}

Each agent assumes that the probability of a volunteer recognising any given
simulated lens as a lens is some number, $\pr(\saidLENS|\LENS,\training)$, that
depends only on what the volunteer is currently looking at, and all the
previous training subjects they have seen (and not on what type of lens it is,
how faint it is, what time it is, \etc). Likewise, it also assumes that the
probability of a volunteer recognising any given dud image as a dud is some
other number, $\pr(\saidNOT|\NOT,\training)$, that also depends only on what the volunteer is currently looking at, and all the
previous training subjects they have seen. These two probabilities define a
2 by 2 ``confusion matrix,'' which the agent updates, every time a
volunteer classifies a training subject, using the following
very simple estimate:
\be
  \pr(``X"|X,\training) \approx \frac{N_{``X"}}{N_X}.
  \label{eq:app:fraction}
\ee
Here, $X$ stands for the true classification of the subject, \ie either
$\LENS$ or $\NOT$, while $``X''$ is the corresponding classification
made by the volunteer on viewing the subject. $N_X$ is the number of
training subjects of the relevant type the volunteer has been shown,
while $N_{``X"}$ is the number of
times the volunteer got their classifications of this type of training subject
right. $\training$ stands for all
$N_{\LENS} + N_{\NOT}$ training data that the agent has heard about to
date.

The full confusion matrix of the $k^{\rm th}$ volunteer's agent is therefore:
\begin{align}
  \CM^k &=
  \begin{bmatrix}
    \pr(\saidLENS|\NOT,\trainingk) & \pr(\saidLENS|\LENS,\trainingk) \\
    \pr(\saidNOT |\NOT,\trainingk) & \pr(\saidNOT |\LENS,\trainingk)
  \end{bmatrix}, \notag \\
        &=
  \begin{bmatrix}
    \CM_{LN} & \CM_{LL} \\
    \CM_{NN} & \CM_{NL}
  \end{bmatrix}^k.
  \label{eq:confmat}
\end{align}
Note that these probabilities are normalized, such that
$\pr(\saidNOT |\NOT) = 1 - \pr(\saidLENS|\NOT)$.

Now, when this volunteer views a test subject,
it is this confusion matrix that will allow their software agent to update the
probability of that test subject being a $\LENS$. Let us suppose that
this subject has never been seen before: the agent assigns a
prior probability that it is (or contains) a lens is
\be
  \pr(\LENS) = p_0
\ee
where we have to assign a value for $p_0$. In the \cfhtls, we might expect
something like 100 lenses in 430,000 images, so $p_0 = 2\times10^{-4}$
is a reasonable estimate. The volunteer then makes a classification $C_k$
($= \saidLENS$ or $\saidNOT$).
We can apply Bayes' Theorem to derive how the agent should
update this prior probability into a posterior one using this new information:
\begin{align}
  \label{eq:app:first}
  & \pr(\LENS|C_k,\trainingk) = \\
  & \frac{\pr(C_k|\LENS,\trainingk)\cdot\pr(\LENS)}
{\left[ \pr(C_k|\LENS,\trainingk)\cdot\pr(\LENS) + \pr(C_k|\NOT,\trainingk)\cdot\pr(\NOT) \right]},
  \notag
\end{align}
which can be evaluated numerically using the elements of the confusion
matrix.

For example, suppose we have a volunteer who is always right about the true
nature of a training subject.
Their agent's confusion matrix would be
\be
  \CM^{\rm perfect} =
  \begin{bmatrix}
    0.0 & 1.0 \\
    1.0 & 0.0
  \end{bmatrix}.
\ee
On being given a fresh subject that actually is a $\LENS$, this hypothetical
volunteer would submit $C = \saidLENS$.  Their agent would then calculate the
posterior probability for the subject being a $LENS$ to be
\begin{align}
  \pr(\LENS|\saidLENS,\trainingk) &= \frac{1.0 \cdot p_0}
           {\left[ 1.0\cdot p_0 + 0.0\cdot(1 - p_0) \right]}
   &= 1.0,
\end{align}
as we might expect for such a {\it perfect} classifier.  Meanwhile, a
hypothetical volunteer who (for some reason) wilfully always submits the wrong
classification would have an agent with the column-swapped confusion matrix
\be
  \CM^{\rm obtuse} =
  \begin{bmatrix}
    1.0 & 0.0 \\
    0.0 & 1.0
  \end{bmatrix},
\ee
and would submit $C = \saidNOT$ for this subject. However, such a volunteer
would nevertheless be submitting useful information, since given the above
confusion matrix, their software agent would calculate
\begin{align}
  \pr(\LENS|\saidNOT,T_k) &= \frac{1.0 \cdot p_0}
           {\left[ 1.0\cdot p_0 + 0.0\cdot(1 - p_0) \right]}
   &= 1.0.
\end{align}
{\it Obtuse} classifiers turn out to be as helpful as {\it perfect} ones, because
the agents know to trust the opposite of their classifications.


\subsection{Online SWAP: Updating the Subject Probabilities}
\label{sec:swap:online}

Suppose the $k+1^{\rm th}$ volunteer now submits a classification, on the same
subject just classified by the $k^{\rm th}$ volunteer. We can generalise
\Eref{eq:app:first} by replacing the prior probability with the current
posterior probability:
\begin{align}
  \label{eq:app:update}
  \pr(\LENS & |C_{k+1},\training_{k+1},\data) = \\
  & \frac{1}{Z} \pr(C_{k+1}|\LENS,\training_{k+1}) \cdot \pr(\LENS|\data) \\ \notag
{\rm where}\;\; Z = & \pr(C_{k+1}|\LENS,\training_{k+1})\cdot\pr(\LENS|\data) \\ \notag
      & + \pr(C_{k+1}|\NOT,\training_{k+1})\cdot\pr(\NOT|\data), \notag
\end{align}
and $\data = \{C_k,\trainingk\}$ is the set of all previous
classifications, and the set of training subjects seen by each of those
volunteers.
$\pr(\LENS|\data)$ is the fundamental property of each test subject that
we are trying to infer. We track $\pr(\LENS|\data)$ as a function of time,
as it is updated;
by comparing it to a lower or upper threshold, we can make decisions about
whether to retire the subject from the classification interface, or
flag it for further study, respectively.

Finally, the confusion matrix obtained from the application of
\Eref{eq:app:fraction} has some inherent noise which reduces as the
number of training subjects classified by the agent's volunteer
increases. For simplicity, the discussion has thus far assumed the case
when the confusion matrix is known perfectly; in practice, we allow for
uncertainty in the agent confusion matrices by averaging over a small
number of samples drawn from Binomial distributions characterised by the
matrix elements $\pr(C_k|\LENS,\trainingk)$ and  $\pr(C_k|\NOT,\trainingk)$. The
associated standard deviation in the estimated subject probability
provides an error bar for this quantity.

In \Sref{sec:results} and \Aref{appendix:swap} below, we define several
quantities based on the probabilities listed above that serve to quantify the
performance of the crowd in terms of the information they provide via their
classifications, and report on the performance of the system in returning a
sample of lens candidates as a function of $\pr(\LENS|C,T)$ threshold.


\subsection{Offline SWAP}
\label{sec:swap:offline}

The probabilistic model described above does not need to be implemented as an
online inference. Indeed, it might be more appropriate to perform the inference
of all agent confusion matrix elements and subject probabilities simultaneously.
Such a joint analysis would implement in full the software agents' basic model assumption
that their volunteers have innate and unchanging talent for lens spotting, that
is parameterised by two constant confusion matrix elements which simply need to
be inferred given the data.

We refer to this alternative calculation as ``offline'' analysis, because it
does not need to be carried out one classification at a time (and hence can be
done at any time after the data is collected). Note that in this offline
inference, the effect will be that of applying the time-averaged confusion
matrices to each classification,  rather than a set that evolves as the agents
(and in the real world, the volunteers) learn. This will mean  that the early
classifications will effectively not be downweighted as a result of the agent's
ignorance. On the other hand, this ignorance provides some conservatism,
reducing the noise due to early classifications if they are unreliable: in
practice we do downweight (via Laplace/add-one smoothing)
the early classifications made by each volunteer, for
this reason. Whether or not an offline analysis out-performs an online one is a
matter for experiment.

The mechanics of how we carry out the offline inference  will be presented
elsewhere (Davis et al, in preparation). Here we briefly note that the procedure
is to maximize the joint posterior probability distribution for all the model
parameters (some 74,000 confusion matrix elements and 430,000 subject
probabilities) with a simple expectation-maximisation algorithm. This takes
approximately the same CPU time as the \StageTwo online analysis, because no
matrix inversions are required in the algorithm. The algorithm scales well, and
is actually faster than the online analysis with the larger \StageOne dataset.
The expectation-maximisation algorithm is robust to initial starting parameters
in, e.g., initial agent confusion matrix elements and Subject probabilities. The
difference in performance between the online and offline analyses is presented
in  \Sref{sec:results:sample} below.


\subsection{Application to the CFHTLS Classifications}
\label{sec:swap:practicalities}

\begin{figure}
\centering\includegraphics[width=0.9\linewidth]{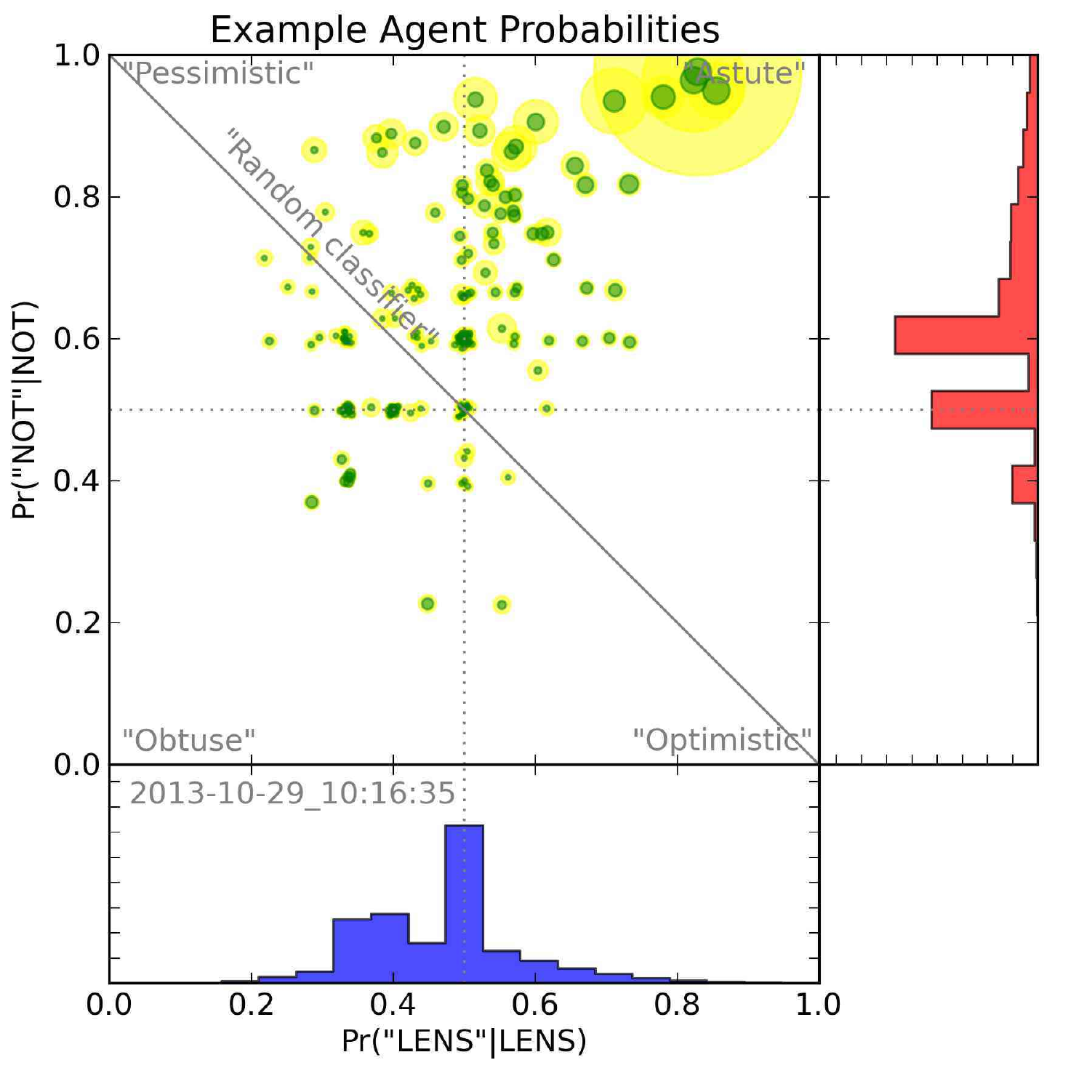}
\caption{Typical \SW agent confusion matrix elements. At a particular
snapshot, 200 randomly-selected agents are shown distributed over the unit plane, with a
tendency to move towards the ``astute'' region in the upper right hand
quadrant as each agent's volunteer views more images. Yellow point size is
proportional to the number of images classified; green point size shows
agent-perceived ``skill'' (\Aref{appendix:swap}).}
\label{fig:swap:agent-probabilities}
\end{figure}

\Fref{fig:swap:agent-probabilities} shows the confusion matrix elements of 200
randomly-selected agents, as they were on a particular day during the \StageOne
online analysis. Many volunteers classify  only a small number of images, and so
their software agents' confusion matrix elements have remained close to their
initial values of (0.5,0.5). As more images are classified (shown by the yellow
point size), the agents' matrix elements tend to move towards higher values,
partly as the volunteers attain greater skill levels (green point sizes, see
\Sref{sec:results} below) but mostly as the agent updates its confusion matrix.
In this upper right-hand quadrant, the agents perceive their volunteers to be
``astute.''

\begin{figure}
\centering\includegraphics[width=0.9\linewidth]{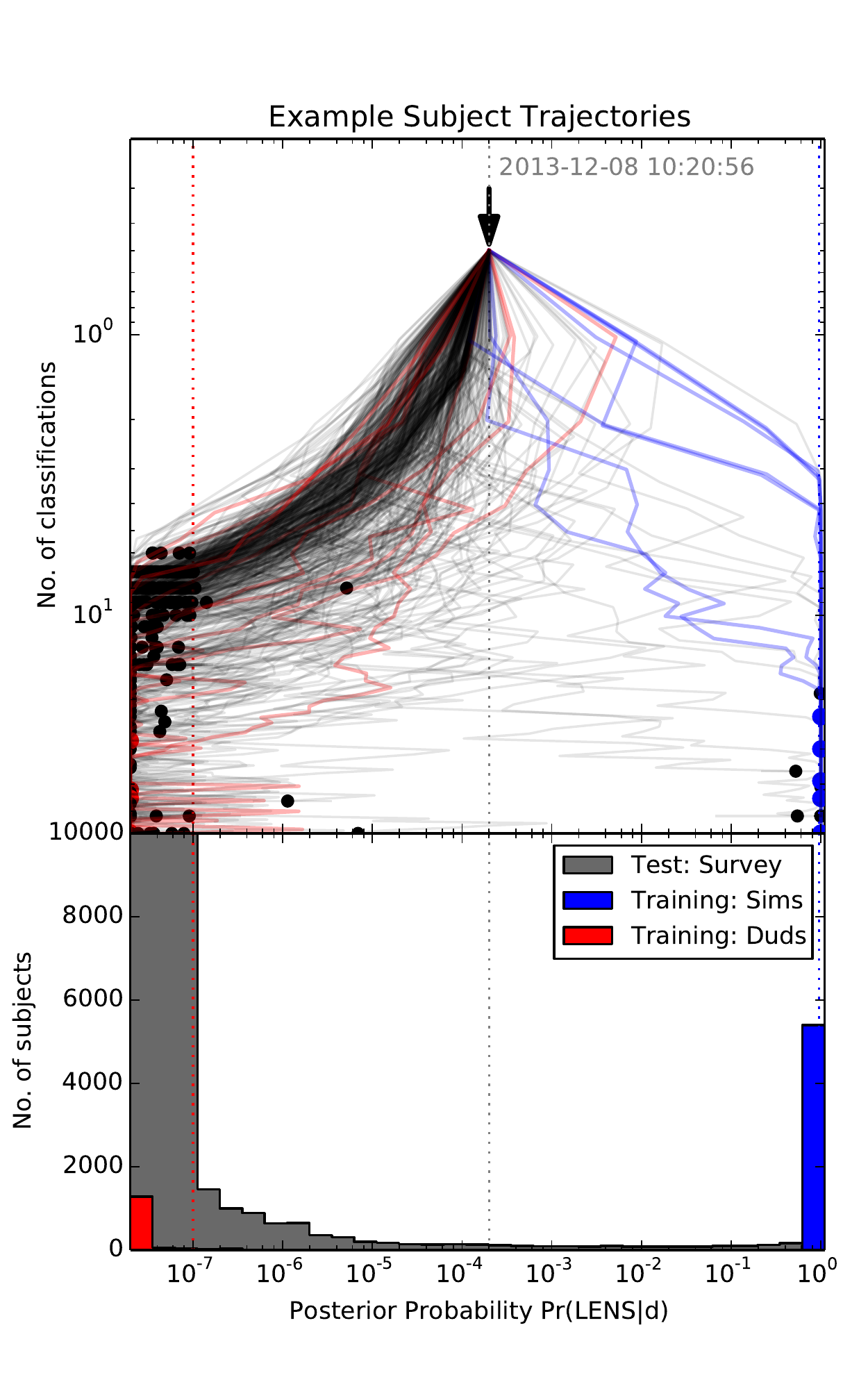}
\caption{Typical \SW \StageOne subject trajectories. Top: 200 randomly-selected
subjects drift downwards as they are classified, while being nudged left and
right in probability by the agents as they interpret the volunteers input. The
dotted vertical lines show (left to right) the retirement threshold, prior
probability, and the detection threshold. Different colours denote the different
kinds of subject. Bottom panel: histograms of all the subject probabilities
computed to date, sub-divided by subject kind. The blue bar (of correctly-detected
sims) hides a grey bar of around 3000 new lens candidates, which are the
subject of \PaperTwo.}
\label{fig:swap:subject-trajectories}
\end{figure}

During the \StageOne classification of the \cfhtls images, we assigned a prior
probability for each image to contain a lens of $2\times10^{-4}$, based on a
rough estimate of the number of expected lenses in the survey, and the fraction
of the survey area covered by each image. We then assigned two values of the
images' posterior probability, $\pr(\LENS|C,T)$, to define ``detection'' and
``rejection'' thresholds. These were set to be 0.95 and (approximately
symmetrically in the logarithm of probability), $10^{-7}$. Subjects that
attained probability of less than the rejection threshold were scheduled for
retirement and subsequently ignored by the analysis code. Subjects crossing the
$P=0.95$ detection threshold were also subsequently ignored, but they were not
retired from the website, just so that more volunteers could see them.

The progress of the subjects during the online analysis is illustrated in
\Fref{fig:swap:subject-trajectories}. Subjects appear on this plot at the tip of
the arrow, at zero classifications and prior probability; they then drift
downwards as they are classified by the crowd, with each agent applying the
appropriate ``kick'' in probability based on what it hears its volunteer say.
Encouragingly, sims (blue) tend to end up with high probability, while duds
(red) pile up at low probability; test subjects (black) mostly drift to low
probability, but some go the other way. The latter will help make up the
candidate sample. As this plot shows, around 10 classifications are typically
required for a subject to reach the retirement (or detection) threshold.

The online analysis code was run every night during the project, and subjects retired
in batches after its completion. This introduced some inefficiency, because
some classifications were accumulated in the time between them crossing the
rejection threshold and the subject actually being retired from the website.
(We quantify this inefficiency in \Sref{sec:discuss:efficiency} below.)
As subjects were retired from the site, more subjects were activated. In this
way, the volunteers who down-voted images for not containing any lensed
features enabled new images to be shown to other members of the community.

When all the subjects had either been retired, or at least
classified around 10 times or
more, the web app was paused and reconfigured for \StageTwo. The sample of
subjects classified during \StageTwo was selected to be all those that passed
the detection threshold ($\pr(\LENS|C,T) > 0.95$) at \StageOne. These were
classified for one week, with no retirement but a maximum classification number
of 50 each. The number of subjects at \StageTwo was small enough that we did not
need to retire any: instead, we simply collected classifications for a fixed
period of time (about 4 weeks). Without the time pressure motivating an
online-only calculation (as there had been during \StageOne), we carried out an
offline analysis (\Sref{sec:swap:offline} above) of the  \StageTwo
classifications as well, both  for comparison and as we will see in the next
section, to improve the pipeline performance.


\begin{figure*}
\begin{minipage}{0.48\linewidth}
  \centering\includegraphics[width=\linewidth]{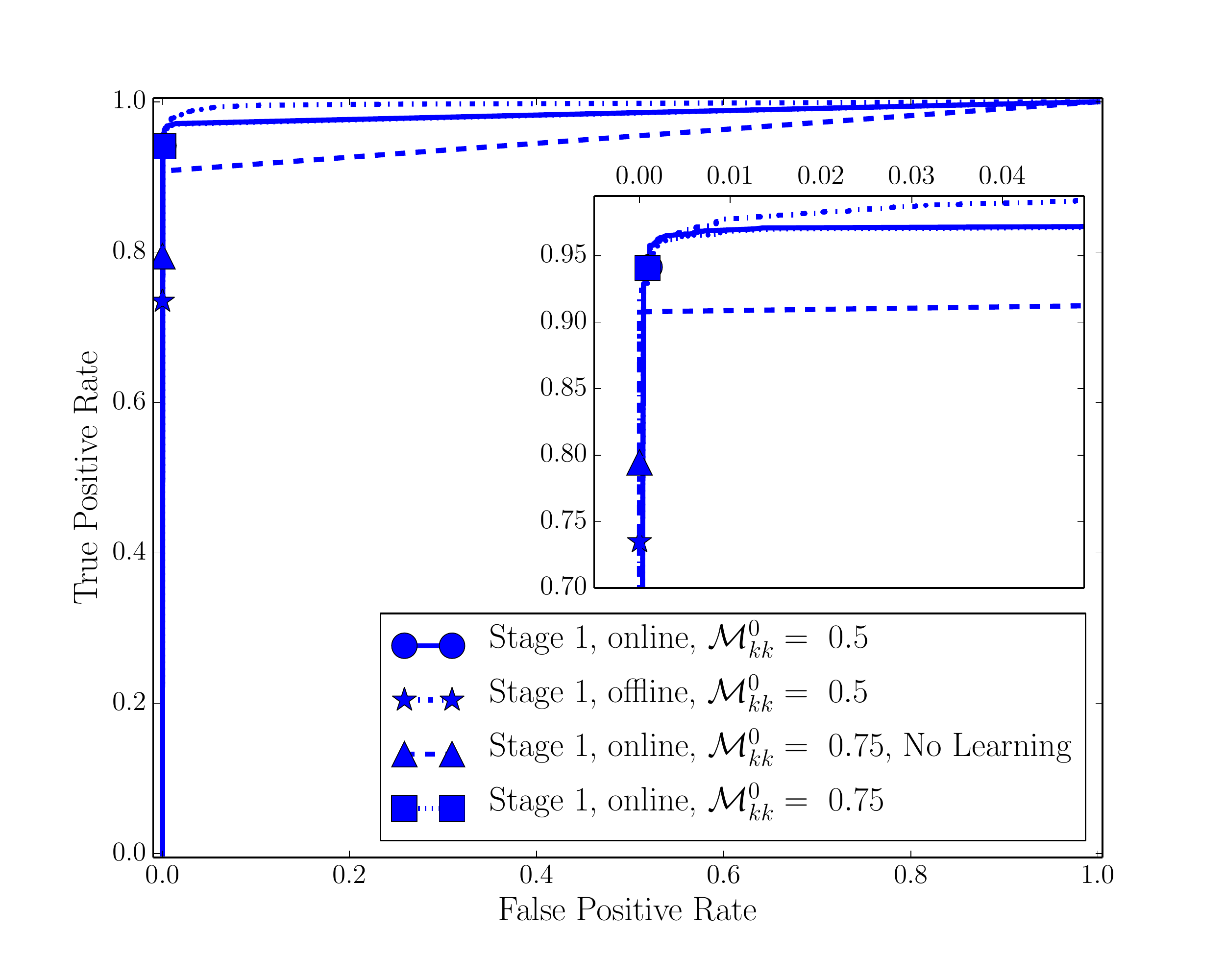}
\end{minipage}\hfill
\begin{minipage}{0.48\linewidth}
  \centering\includegraphics[width=\linewidth]{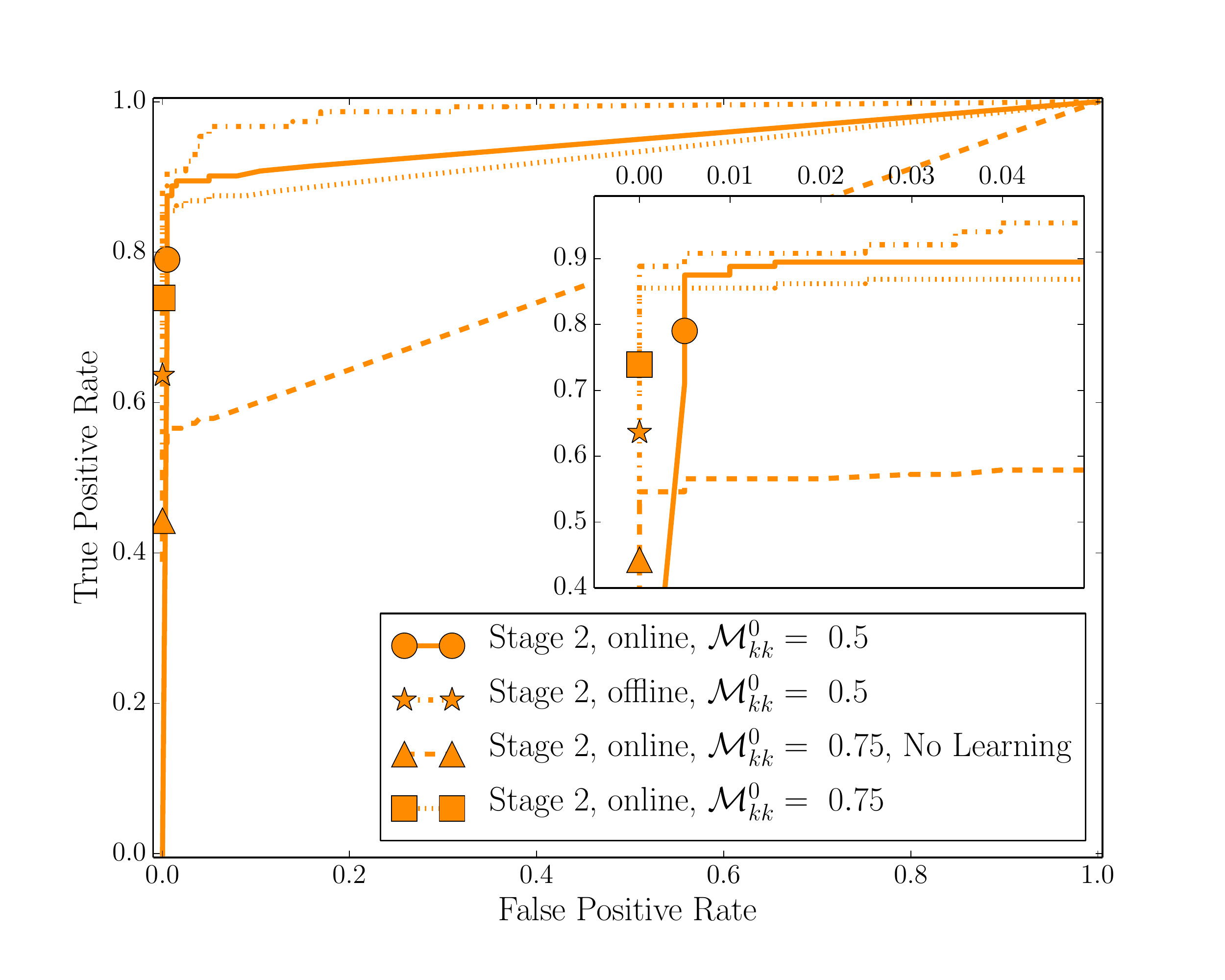}
\end{minipage}
\caption{Receiver operating characteristic curves for the \SW system, using
the \cfhtls training set. Left: \StageOne, right: \StageTwo. Insets show a
magnified view of the top lefthand corner of each plot. Linestyles illustrate
the software agent properties (initial confusion matrix elements, learning or not)
discussed in the text, as well as the difference between the online and offline
analyses. The \StageTwo sample was defined using the online \StageOne results
(solid blue curve), while the offline analysis was chosen for the final
\StageTwo results (dot-dashed orange curve).
Lens probability threshold varies along the curves: the symbols
indicate the points where
the value of this quantity is 0.95. In the lefthand panel, the circle
point is masked by the square point.}
\label{fig:results:sample:roc}
\end{figure*}

\section{Results}
\label{sec:results}

In this section we present our findings about the performance of the \SW system,
in terms of the classification of the training set, the  information contributed
by the crowd, and the speed at which the image set was classified.


\subsection{Sample Properties}
\label{sec:results:sample}

We first quantify the performance of the \SW system in terms of the recovery of
the training set images. At \StageOne, this set contained around 5712 simulated
lenses, and 450 duds; at \StageTwo, we used 152 images of simulated lenses and
201 duds selected as \StageOne false positives (\Sref{sec:design:stages:two}).
\Fref{fig:results:sample:roc} shows receiver operating characteristic (ROC)
curves for \cfhtls \StageOne and \StageTwo. These plots show the true positive
rate (TPR, the number of sims correctly detected divided by the total number of
sims in the training set), and the false positive rate (FPR, the number of duds
incorrectly detected divided by the total number of duds in the training set), both
for a given sample of detections defined by a particular probability threshold,
which varies along the curves.  In both stages, these curves show that true
positive rates of  around 90\% were achieved, at very low false positive rates.
The probability threshold that was used for retirement during the \StageOne online analysis
was 0.95; this point is marked with a symbol on each curve.
This turned out to be close to optimal (although a better approach would be to
keep track of the ROC curve as the survey progressed!).

For comparison we show the results of an analysis where the classifications of
training images were ignored, and the agents' confusion matrix elements
were instead all simply assigned initial values of 0.75,
which then remained constant. This set-up emulates a
very simple unweighted voting scheme, where all classifications are treated
equally.  In this case, the TPR never reaches 80\% in \StageOne or 60\% in
\StageTwo, thus illustrating the benefit of including training images and
allowing the agents to learn via their Bayesian updates.
When the software agents are allowed to update their confusion matrices,
the choice of
initial confusion matrix is not very important: the same 0.75 initial values
applied to normal, learning agents resulted in only a slightly lower TPR than
the default case.

The dot-dashed curves show the results from the offline analysis. At \StageOne the
results are very similar to the online version that was actually run (solid
line). However, at \StageTwo there is marginal evidence of there being greater
benefit to doing the analysis offline (\Sref{sec:swap:offline}).
Over 85\% TPR is achieved at zero FPR in the offline analysis , while if one is
willing to accept a false positive rate of 5\%, the true positive rate rises to
over 95\%, showing that some of the sims that were missed in the online analysis
may be being recovered by doing the analysis offline. (The same is true at
\StageOne, but to a lesser extent.) One interpretation for this result (if it holds up)
would be that the offline analysis, by using each agent's entire history, is
making less noisy probability estimates. This could be consistent with
other citizen science/crowdsourcing projects where,
in the absence of high quality information about classifiers,
sophisticated strategies tend to under-perform naive but simpler
ones \citep[][although we note that this rule of thumb seems not to extend
to simple voting in this case!]{Waterhouse}.

Assuming Poisson statistics for the
fluctuations in the numbers of recovered lenses, the uncertainty in the measured
\StageTwo TPR values is around 8\%, but the online and offline samples are
highly correlated, such that the uncertainty on the difference between the ROC
curves is somewhat less than this.
Still, a larger validation set is needed to
test these algorithmic choices more rigorously. At \StageOne, high TPR can be
measured to better than 1\%.

\begin{figure}
\centering\includegraphics[width=\linewidth]{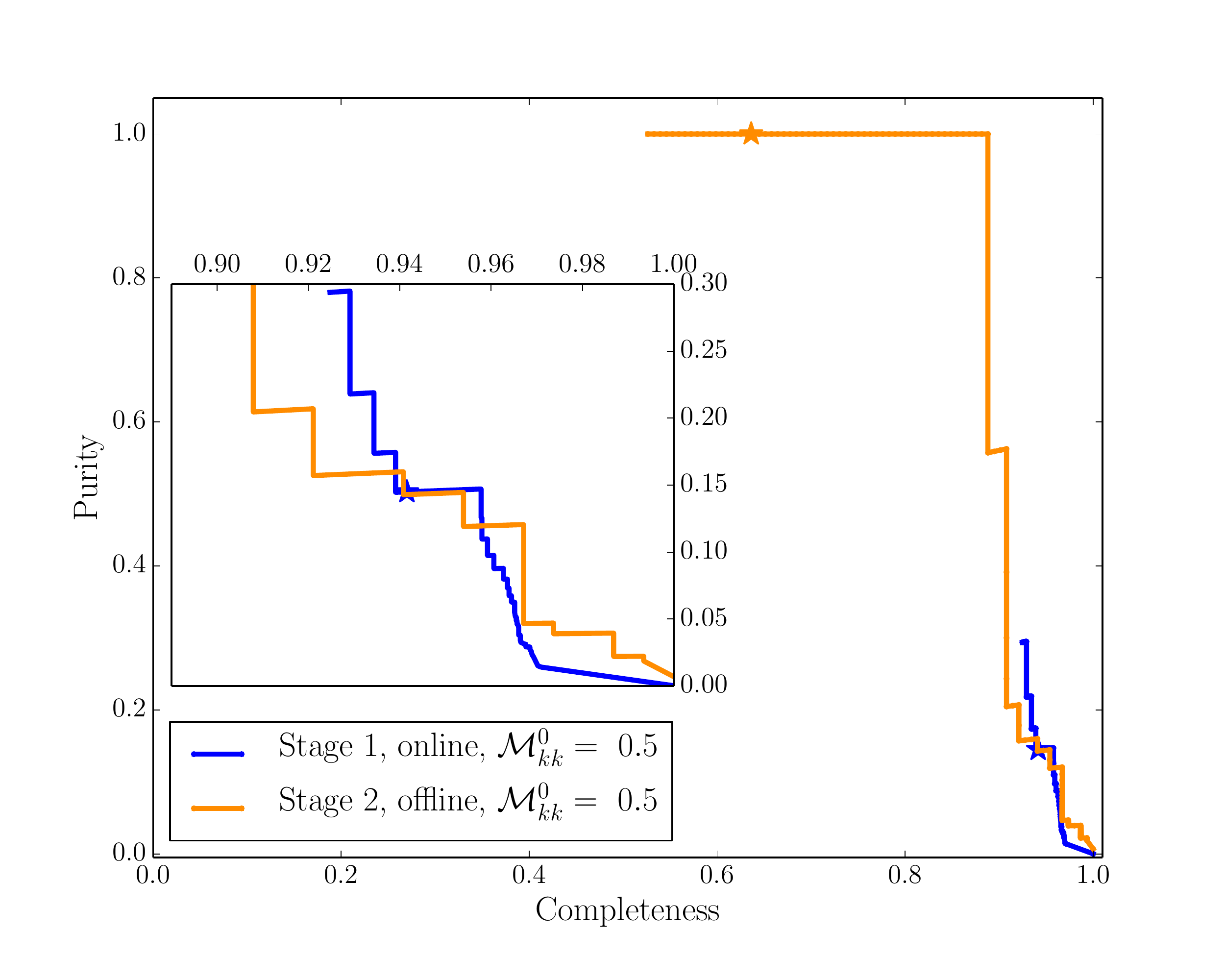}
\caption{Completeness-estimated purity curves for the \sw system, using
the \cfhtls training set. The curves are truncated at the high purity end by
the ``detection'' limit subject probability (subjects in the online
\StageOne analysis were
deemed to be detected at $p=0.95$, while at \StageTwo rounding error led to
an upper limit of $p=1.0$). The inset shows a magnified view of the
bottom righthand corner of the plot.}
\label{fig:results:sample:CP}
\end{figure}

Adopting the online \StageOne analysis, and the offline \StageTwo analysis, we
show in \Fref{fig:results:sample:CP} a plot of the more familiar (to
astronomers) quantities of completeness versus purity, again for
the two stages. As in \Fref{fig:results:sample:roc}, the detection threshold
varies along the curves. Completeness is defined as the number of correctly
detected sims divided by the total number of sims in the training set, while
purity is the number of correct detections divided by the total number of
detections.\footnote{The completeness is equivalent to the TPR and is also
known elsewhere as the ``recall.''  The purity is also known as the
``precision.''}
If the training set were to be sampled fairly from test set, the
completeness of the training set would be equal to the completeness of
the test set. (In practice this will likely only be approximately
true, as simulated lenses [and the distributions of their properties]
are used instead of real ones.)

The purity depends on the proportion of sims to duds, and so the
purity of the test set must be approximated by rescaling the training
set to the expected proportion of lens systems to not-lens systems in
the survey. We expect there to be around 90 lenses in the CFHTLS
already (a rate of 1 lens in every 5000 images or so); to a very good
approximation  the number of non-lens images in the survey is just the
number of images in  the survey (some 430,000). First we compute the
expected number of false positives by multiplying the FPR by the
expected number of non-lenses in the survey (430,000).  Then we
multiply the TPR by the expected number of lenses in the survey (90),
to get the expected number of true positives. The sum of the true
positives and the false positives gives the expected sample size;
dividing the expected number of true positives by this sample size
gives the purity. Note that the completeness is invariant to this
transformation. The \StageOne curves are truncated by the retirement
of subjects in this phase, which sets the minimum size of this sample.
We see from the solid blue curve that over 90\% completeness was
able to be reached, albeit in samples with not more than 30\% purity.
We set the detection threshold for \StageOne to be 0.95 (shown by the blue star
in \Fref{fig:results:sample:CP}), leading to a sample with 94\% completeness
and 15\% purity.

Does this performance level vary between the different types of gravitational
lens? To investigate the completeness to the three different types of lens in
the training set, we repeated the same procedure but now considering only the
detections of a certain kind of lens and of the non-lenses in the training set.
We estimate the expected number of lenses and non-lens false positives by
dividing the lens and dud sets into equal fractions. The lensed quasar part of
the training set yielded the highest completeness, suggesting that these were
the easiest sims to spot. The lensed galaxies were recovered at the lowest
completeness, likely due to the difficulty  of separating the lens and source
galaxy light.

At \StageTwo, where no retirement was carried out, it was possible to reach
100\% purity: the knee of the curve is at just under 90\% completeness. However,
the purity decreases rapidly if higher completeness than this is sought. The
optimal sample in this simulated lens search experiment would have been
constructed with a threshold value of $\pr(\LENS|C,T) > 0.47$. At 100\% purity
and 90\% completeness, it would have contained around 89 lens candidates.
However, we remind the reader that all these values are dependent on the
properties of the training subjects, which were chosen to be fairly
visible: we expect the completeness to real lenses to be somewhat lower, as a
result (see \PaperTwo).

Finally, it is worth noting the implications of the results in this
section for future studies of samples of lenses (and lens candidates)
discovered through visual inspection at \SW. The ROC curve analysis we
have carried out should be very familiar to those working in machine
classification, and in fact is identical to that which would be
performed on the classifications made by a new automated method.
Supervised machine learning methods and the \SW system as described
here both require a training set, and both return quantitative,
probabilistic classifications (consisting of a ``label'' and some measure
of confidence in that label) that can be used in further analysis. A
good example of where such quantitification is important is in the
derivation of the selection function for a given sample. While such a
derivation is beyond the scope of this work, it is sufficient to note
that from the point of view of an astronomer seeking to derive a
selection function, {\it the labels produced by \SW and those produced
by a machine learning method can be treated equivalently}: that is, we
have succeeded in elevating visual inspection to a quantitative
science. Indeed, the discussion of completeness given above is all in
terms of recovery of a large set of simulated lenses, just as it would
be in the case of an automated method---and in both cases, the
limiting factor is the realism of the training set. In \PaperTwo we
investigate this limit further by assessing the performance of the \SW
system against the (small) set of real lens candidates known to lie in
the field; for now, we note that the training set used in this work
constitutes a valid sample of objects for any alternative machine
learning lens detection method to be tested against. In future, we
could employ a larger training set, to enable the
selection efficiency to be characterised as a function of multiple
observables. This will be achievable, since
the small training set we used in the current study
was classified around 20 times more
than  was the test set: we could therefore collect many fewer
classifications per training image in a much larger training set.


\subsection{Crowd Properties}
\label{sec:results:crowd}

\begin{figure*}
\centering\includegraphics[width=0.9\linewidth]{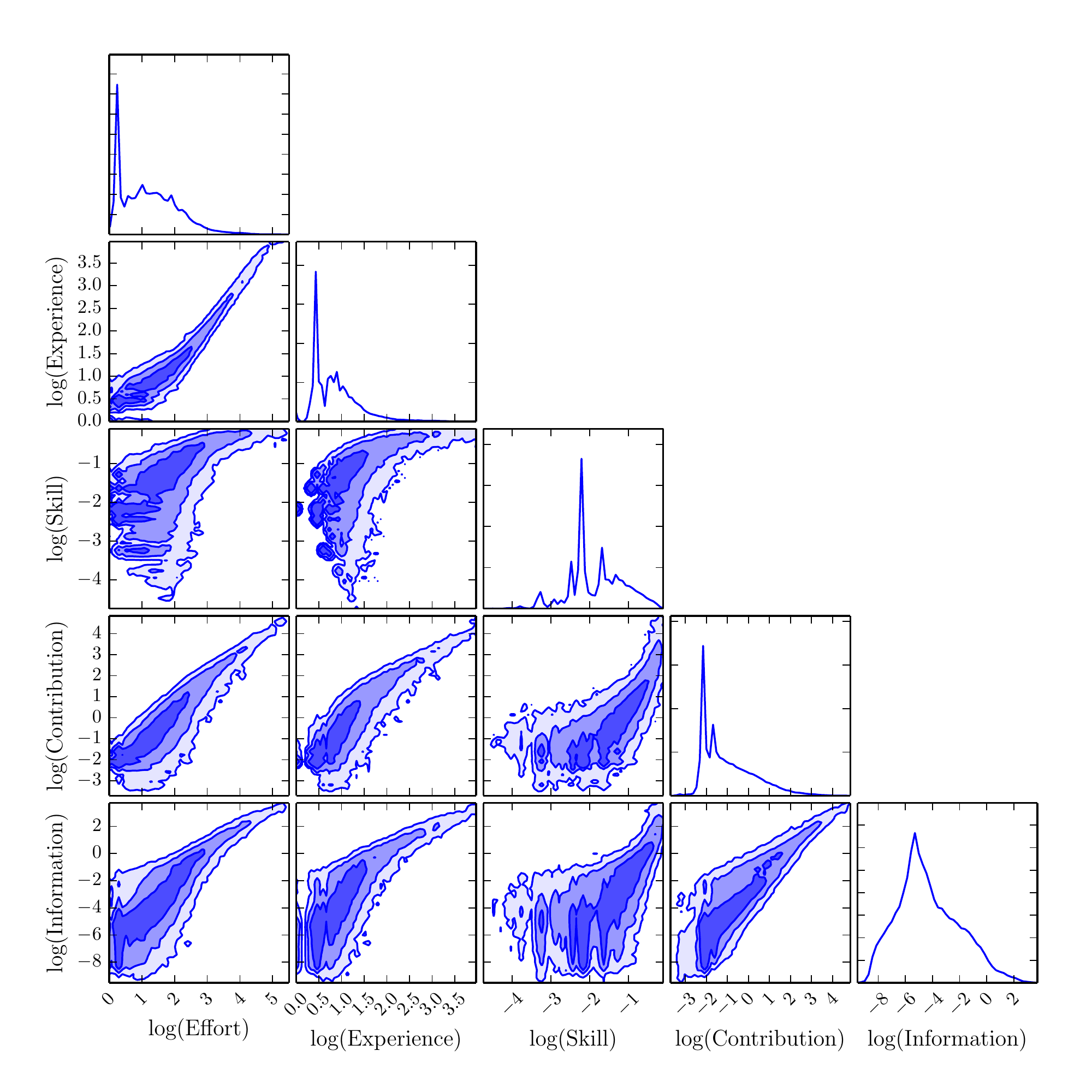}
\caption{Key properties and contributions of the \SW \StageOne crowd.
Plotted are the 1-D
and 2-D smoothed marginalized distributions for  the logarithms of the
properties of the agents described in the text. The contours
contain 68\%, 95\% and 99\% of the distributions.}
\label{fig:crowd:cornerplot}
\end{figure*}

To investigate the properties of the \sw crowd, as characterised by their
agents, we define the following quantities and plot their one and 2-D marginal
distributions in
\Fref{fig:crowd:cornerplot}. This figure only shows the \StageOne agents for clarity,
but the trends we found the same trends in \StageTwo.
\begin{description}
\item{\noindent\bf ``Effort:''} The number of test images, $\effort$, classified
by a volunteer. In \StageOne, the mean effort per agent was 263; in the shorter
\StageTwo it was 81.
\item{\noindent\bf ``Experience:''} The number of training images, $\experience$,
classified by a volunteer. In \StageOne, the mean experience per agent was 29;
in \StageTwo (where the training image frequency was set higher) it was 34.
\item{\noindent\bf ``Skill:''} The expectation value of the information gained
per classification (in bits) by that volunteer,  $\skill$, for subjects which
have lens probability 0.5 (\Aref{appendix:swap}). Random classifiers have
$\skill = 0.0$~bits, while perfect classifiers have $\skill = 1.0$~bit.
All software agents start
with $\skill = 0.0$~bits. The skill of  an agent increases as training subjects are
classified, and the agent's estimates of its confusion matrix elements improve.
In \StageOne, the mean skill per agent was 0.04 bits; in  \StageTwo it was
0.05~bits.
\item{\noindent\bf ``Contribution:''} This is the integrated skill over a volunteer's
{\it test subject} classification history, a quantity representative of the
total contribution to the project made by that volunteer (see \Aref{appendix:swap} for
more discussion of this quantity). The classifications of training images
allow us to estimate the skill, while the classification of test images determines
contribution.  In \StageOne, the mean  contribution per agent was
34.9 bits; in \StageTwo it was 33.5.
\item{\noindent\bf ``Information:''} The total information $\information$
generated by the software agent during the volunteer's classification activity. This
quantity depends on the value of each subject's lens probability when that
subject was presented to the volunteer (simply because information gain is
defined in terms of posterior relative to prior probabilities,
\Aref{appendix:swap}).  As a result, we expect this raw quantity to be noisier
than the ``contribution'' defined above. The lower righthand panel of
\Fref{fig:crowd:cornerplot} confirms this: there is a strong, linear correlation
between contribution and information gain, but at a given contribution, the
distribution of information gain has a tail at high values, corresponding to
volunteers that were lucky in the number of high probability subjects that they
happened to be shown (the information gain per classification is maximized when the
subject probability is 0.5, while most subjects have probabilities closer to or less than the prior).
In most of the following discussion we use the contribution when characterizing
crowd participation, but we include
the information gain for completeness.
\end{description}

The leftmost column of \Fref{fig:crowd:cornerplot} shows how the last four of
these properties depends on the effort expended by the volunteers. As expected,
we see that experience is tightly correlated with effort, reflecting the design
of training images being presented throughout each stage (albeit at decreasing
frequency). We also see a strong correlation between effort and skill, which was
hoped for but not guaranteed: the more images the volunteers see, the better
able to contribute information they are.

The effort distribution shows two peaks, suggestive of two types of
participation.  The sharp spike at just a few images classified presumably
corresponds to visitors who only classify a few subjects before leaving the site
again. The broader hump contains people doing tens to hundreds of
classifications: the skill vs.\ experience panel of \Fref{fig:crowd:cornerplot}
shows this group to achieve a broad distribution of  skill, peaking at around
0.05 but with a long tail to higher values.

At high values of experience and effort, the skill is \emph{always high}. There
seem to be very few agents logging large numbers of classifications at low
skill: {\it almost all high effort
``super-users'' have high skill.} These two properties are reflected in both the
contributions these volunteers make (third row) and the information they
generate (fourth row), and we suggest that this would be a useful metric for
determining ``well-designed'' implementations of citizen science projects.

We found that the distributions for the \StageTwo agents to be qualitatively
very similar to those for the \StageOne agents. The differences are: 1) the
maximum effort possible at \StageTwo is smaller, simply because fewer subjects
were available to be classified, and 2) the information generated per agent was
slightly higher at \StageTwo, just because the subjects had (on average) higher
probability.

\begin{figure}
\centering\includegraphics[width=0.9\linewidth]{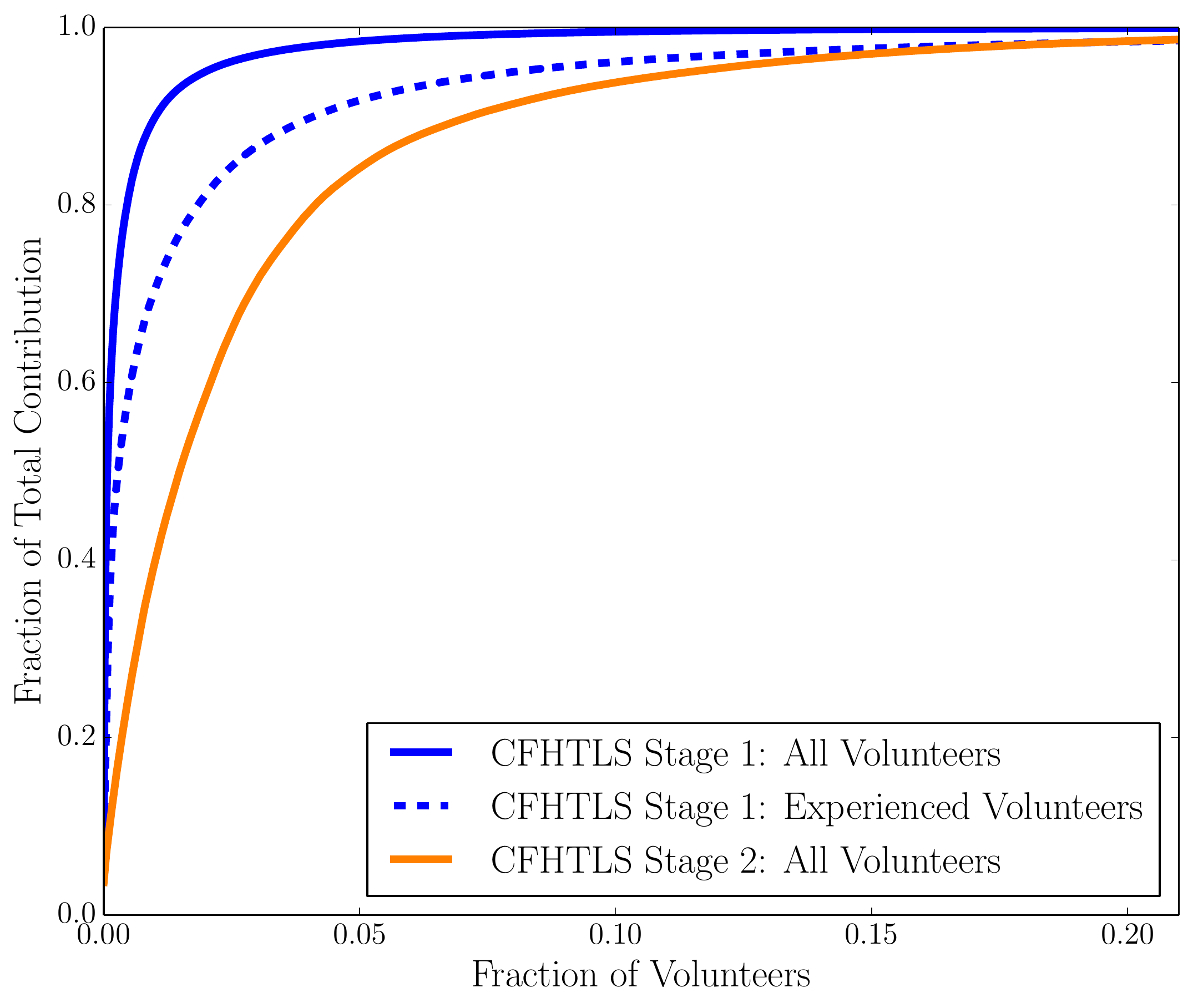}
\caption{Cumulative distributions of the contributions made by the agents: 90\%
of the contribution was made by the highest contributing 1\% of the crowd at
\StageOne (blue), and the highest contributing  7\% of the crowd at \StageTwo
(orange). The \StageOne agents are shown in blue, the \StageTwo agents in
orange. ``Experienced volunteers'' classified 10 or more training subjects.}
\label{fig:crowd:cumulplot}
\end{figure}

\Fref{fig:crowd:cornerplot} shows the \SW crowd to have quite broad
distributions of logarithmic effort, skill, and contribution. To better
quantify the contributions made by the volunteers, we show their cumulative
distribution on a linear scale in \Fref{fig:crowd:cumulplot}. This plot shows
clearly the importance of the most active volunteers: at
\StageOne,  1.0\% of the volunteers -- 375 people -- made 90\% of the
contribution.  At \StageTwo, where it was not possible to make as
many classifications before running out of subjects, 7.2\% of the volunteers
-- 141 people -- made 90\% of the contribution.

\begin{figure}
\centering\includegraphics[width=0.9\linewidth]{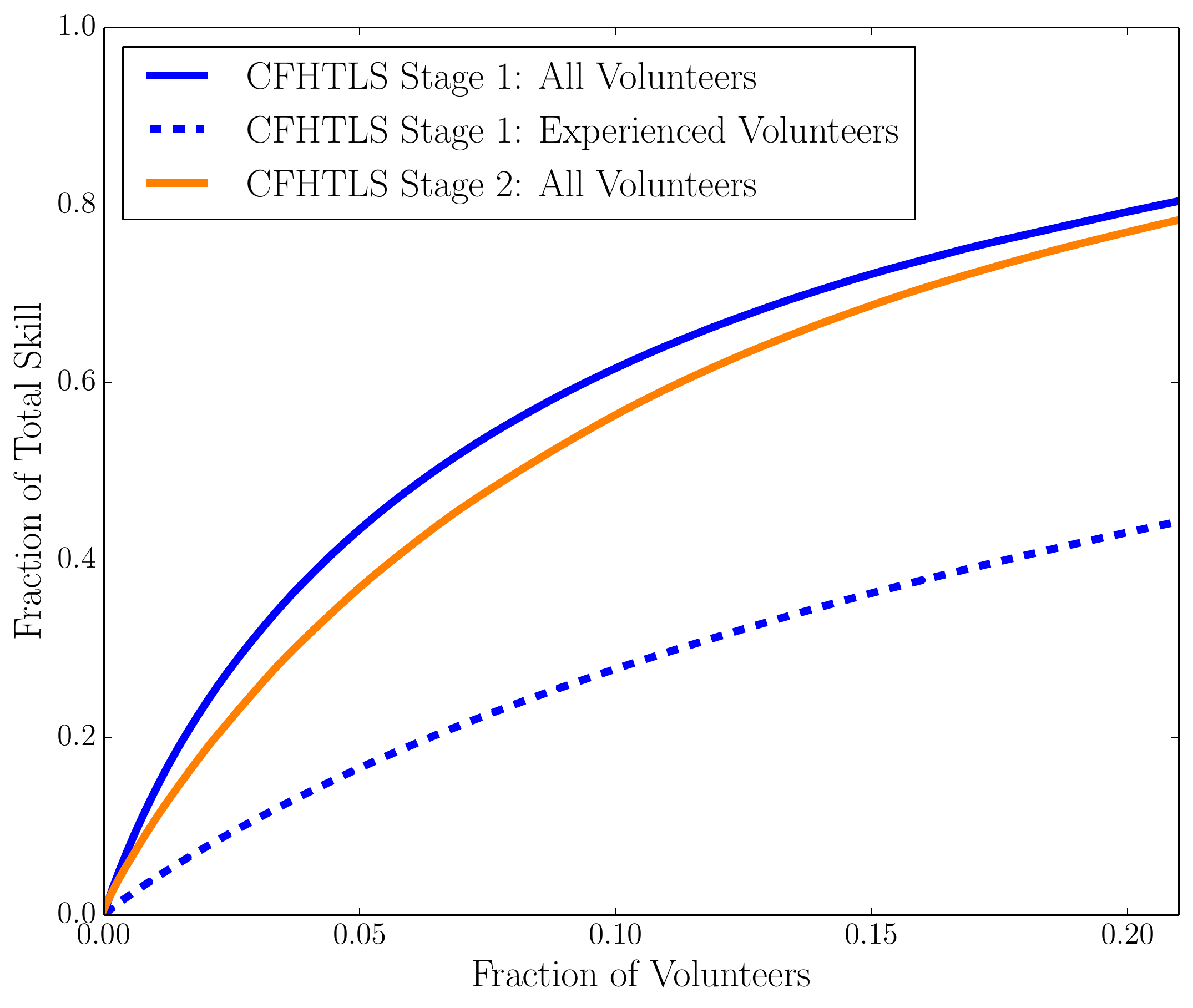}
\caption{Broad cumulative distributions of agent skill:
the most skilled 20\% of the crowd only possess around 80\% of the skill,
at \StageOne. The \StageOne agents
are shown in blue, the \StageTwo agents in orange. ``Experienced volunteers''
classified 10 or more training subjects.}
\label{fig:crowd:cumulskillplot}
\end{figure}

However, it is not the case that only these small groups were {\it capable} of
making large contributions. The cumulative distribution of agent skill is shown
in \Fref{fig:crowd:cumulskillplot}: these distributions are significantly
broader than the corresponding distributions of agent contribution, in
\Fref{fig:crowd:cumulplot}. For example, 80\% of the skill is distributed among
20\% of the agents.
The inexperienced volunteers also possess a significant fraction of the skill.
We find that dividing the crowd into ``experienced volunteers'' (who have seen
10 or more training images), and ``inexperienced volunteers'' (the rest),
results in two groups containing approximately equal total skill (see the dashed
curve in \Fref{fig:crowd:cumulskillplot}): the most skillful 20\% of {\it
experienced} volunteers (1824 people) only possess 43\% of the total skill.
The breadth of the skill distributions suggests that  the high level of
contribution made at \SW by experienced volunteers is largely a matter of choice
(or perhaps, availability of time!). There were many other volunteers that  were
skillful enough to make large contributions -- they just didn't classify as many
images.

\subsection{Classification Speed}
\label{sec:results:speed}

\begin{figure}
\centering\includegraphics[width=\linewidth]{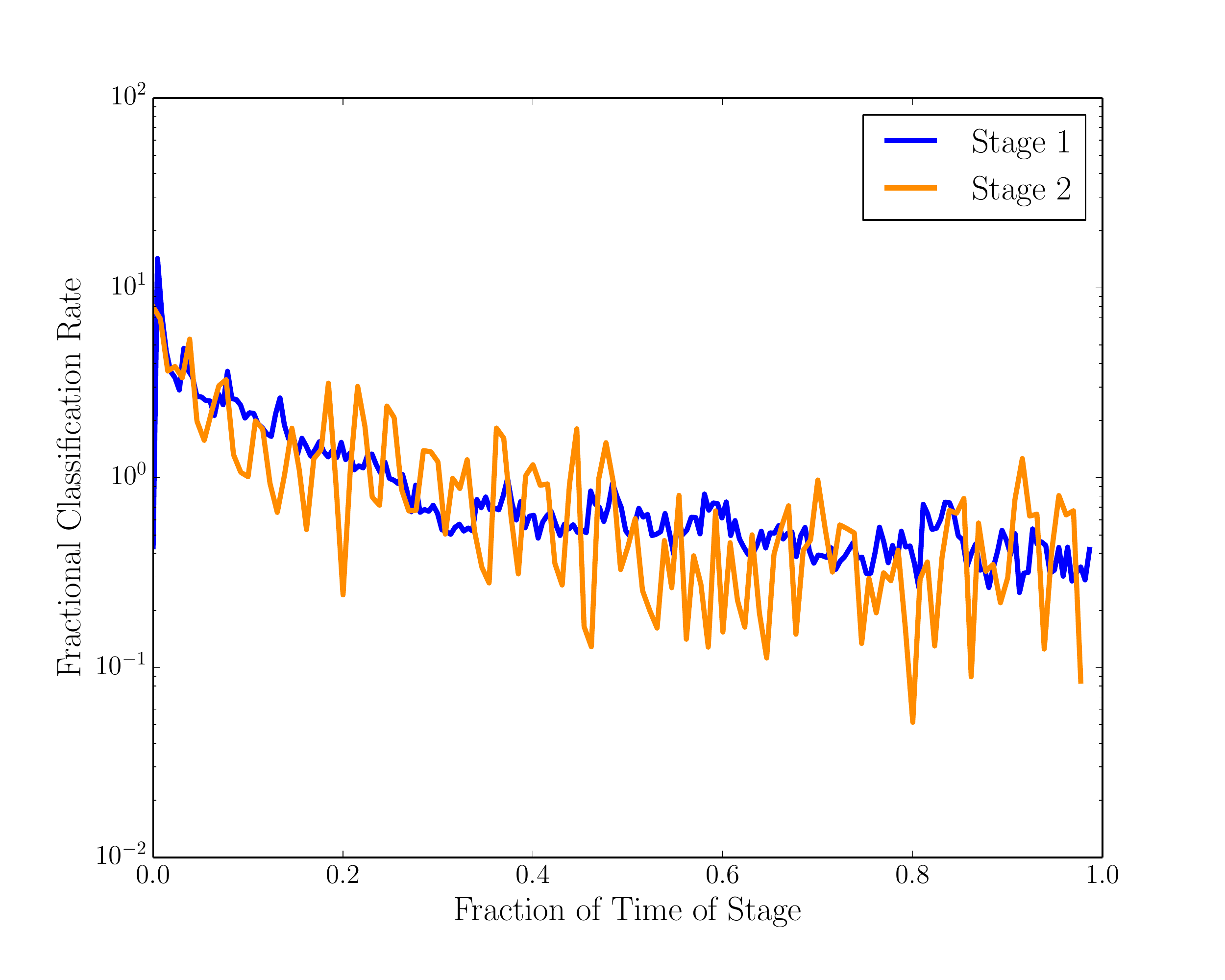}
\caption{Fractional classification rate in the \SW$-$\cfhtls survey. Fractional classification rate is the number
of classifications per unit time, divided by the final total number of
classifications. Fractional survey time is the time elapsed since the beginning
of the survey, divided by the total length of the survey. The unit of time in
the fractional  classification rate is one survey, so that the curves in
integrals under the curves are both one. }
\label{fig:results:speed}
\end{figure}

How fast does the \SW crowd classify subjects? Each software agent records the timestamp
of each classification its volunteer makes; by measuring the time lags between
successive classifications, we can make estimates of the crowd's classification
and contribution speed. We plot the former quantity for both \StageOne and
\StageTwo in \Fref{fig:results:speed}, normalizing the speed and time axes to
their respective totals. The fractional classification rates
in each stage of the survey fall off in approximately the same way, despite the
factor of nearly 20 difference in crowd size. \StageOne (consisting of
$\sim430,000$ subjects) was completed in around 5000 hours (by some 37,000
participants), with classification rates in the first few days averaging at
$\sim10^4$ per hour. This was stimulated by various forms of advertising (press
releases, and emails to registered Zooniverse users). The asymptotic
classification rate was  around ten times lower, at $\sim10^3$ per hour.
Interestingly, the average skill was found to be approximately constant over the
lifetime of \StageOne, leading to a contribution rate that tracks closely the
classification rate. (A 10\% increase in average skill was seen over the first
40 days, and a decrease of the same amount in the last 40 days -- not enough to
cause a significant difference between classification and contribution rate
behaviour, but perhaps reflecting a learning period at the start and then a
decrease
in participation later on.)
\StageTwo (nearly 3400 subjects) was completed in around 600
hours by around 2000 participants, with classification rates starting at
$~\sim2000$ per hour and decaying to $\sim100$ per hour.

The similarities between \StageOne and 2 regarding the results above,
and despite the difference in the task set,
suggest that these numbers can be scaled to estimate cautiously
the speed of future \SW projects. For example, the completion time for a \SW
project may be approximated as
\begin{equation}
    \tau \approx \tau_0 \left(\frac{10^4}{N_p}\right)
                               \left(\frac{N_s}{10^4}\right),
\label{eq:speed}
\end{equation}
where $N_s$ is the number of subjects to be classified, and $N_p$ is the number
of volunteers in the crowd. Just using the numbers above, we find a characteristic
timescale of $\tau_0 = 18\,{\rm days}$, but as we will see in the next section,
we expect this to be significantly shorter in future projects.


\section{Discussion}
\label{sec:discuss}

What can we learn from the results of the previous section, for future
projects? Potential improvements to the \SW system can be divided into three
categories: performance, efficiency and capacity.


\begin{figure*}
\begin{minipage}{\linewidth}
  \begin{minipage}[t]{0.47\linewidth}
    \begin{minipage}{0.46\linewidth}
      \centering\includegraphics[width=\linewidth]{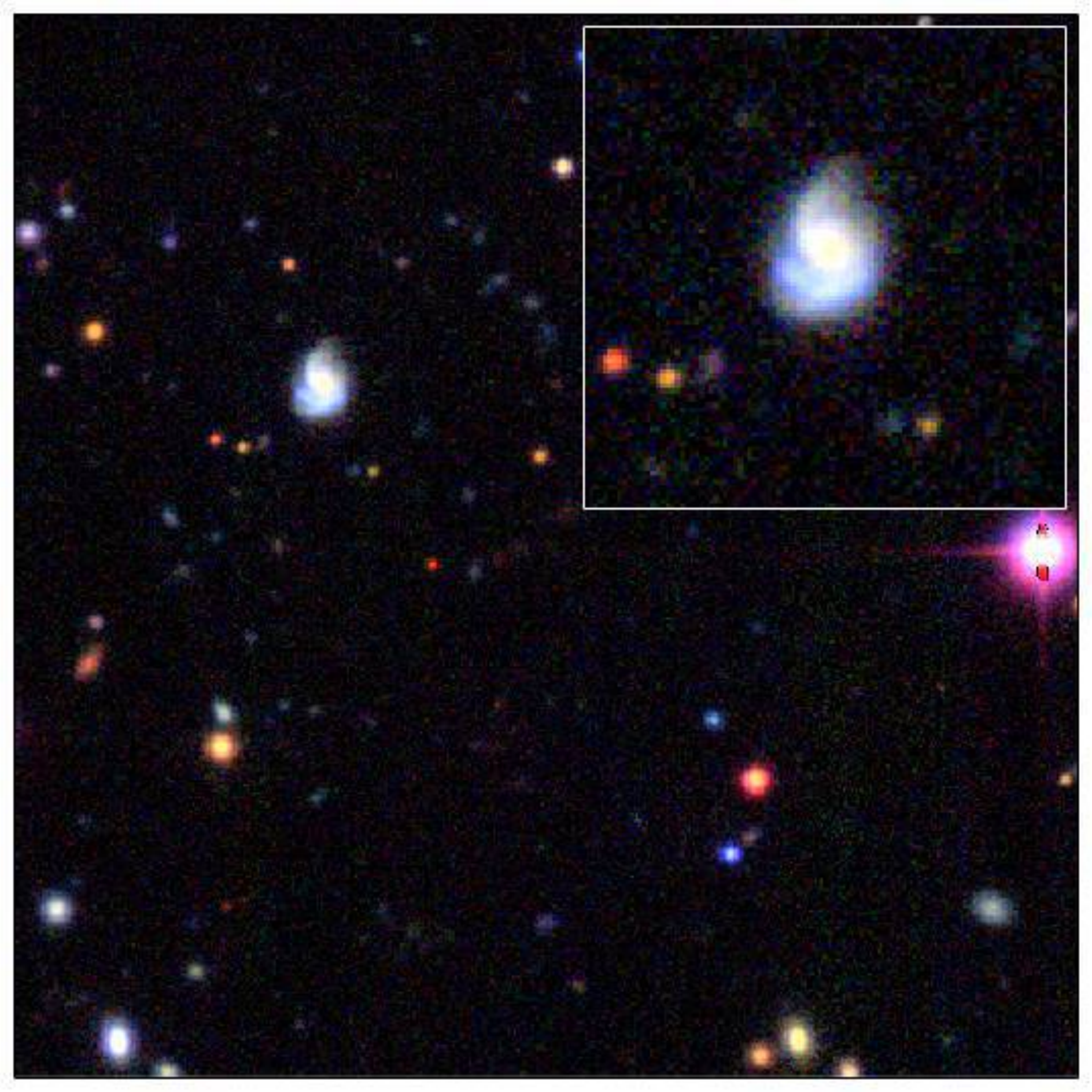}
    \end{minipage}\hfill
    \begin{minipage}{0.50\linewidth}
      \centering\includegraphics[width=\linewidth]{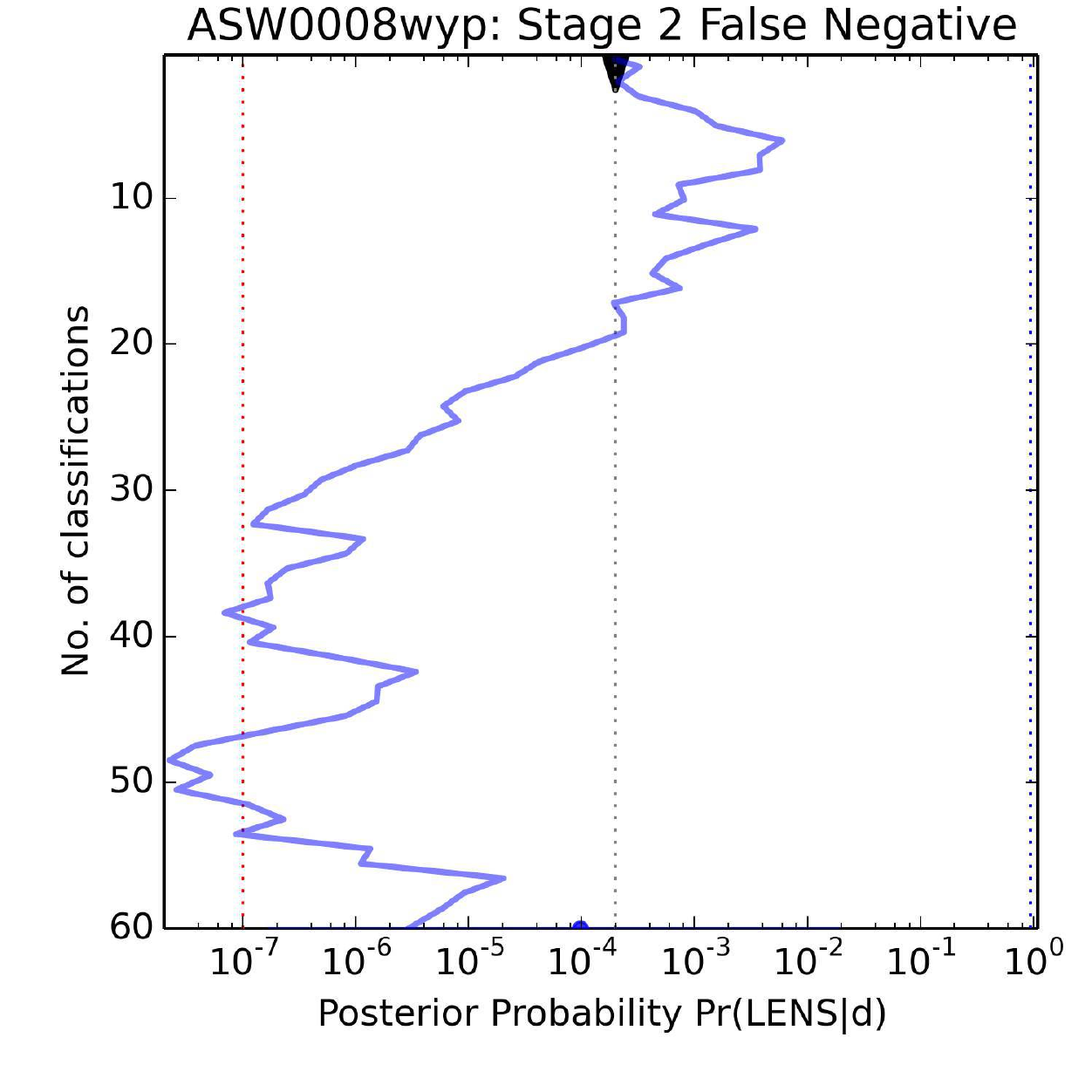}
    \end{minipage}
  \end{minipage}\hfill
  \begin{minipage}[t]{0.47\linewidth}
    \begin{minipage}{0.46\linewidth}
      \centering\includegraphics[width=\linewidth]{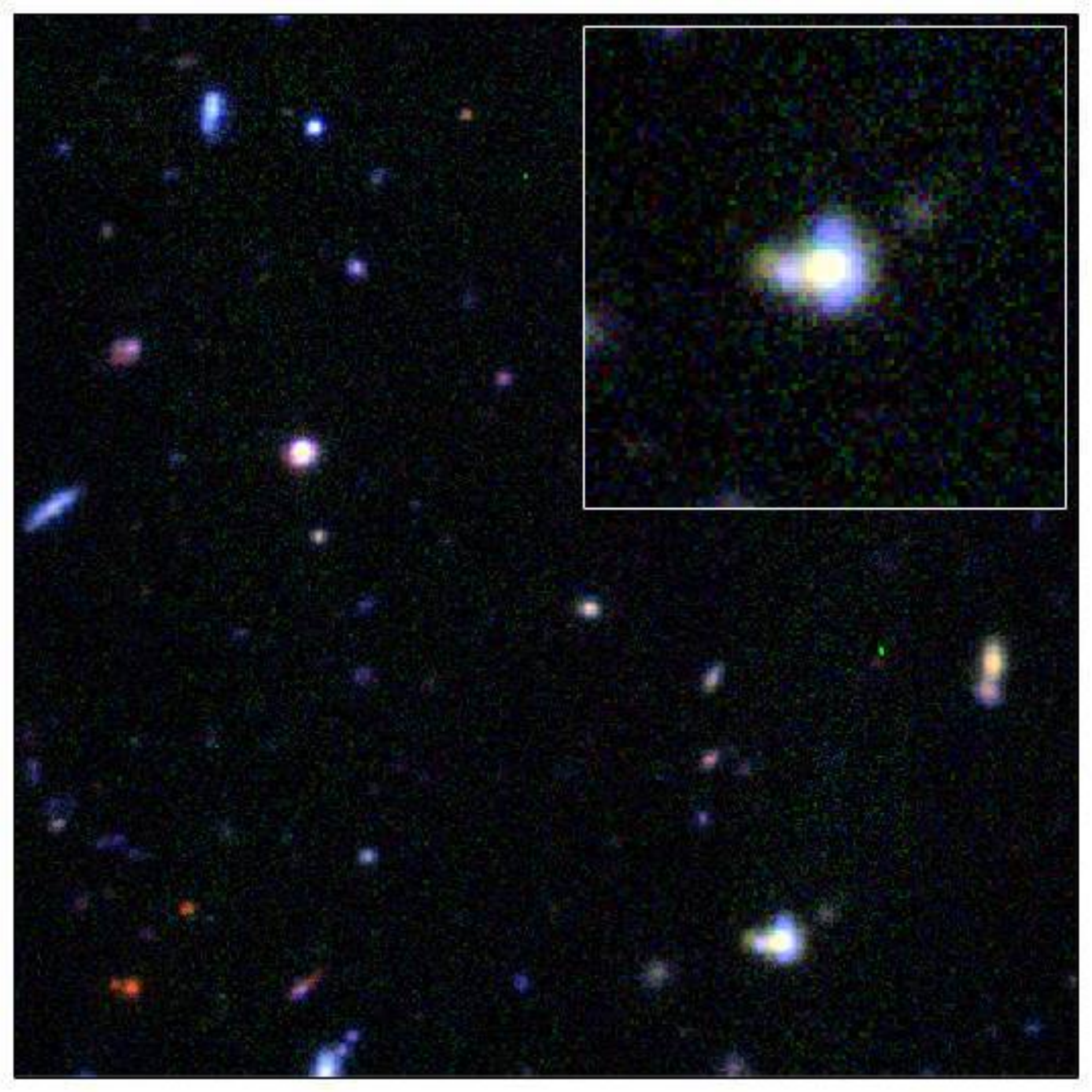}
    \end{minipage}\hfill
    \begin{minipage}{0.50\linewidth}
      \centering\includegraphics[width=\linewidth]{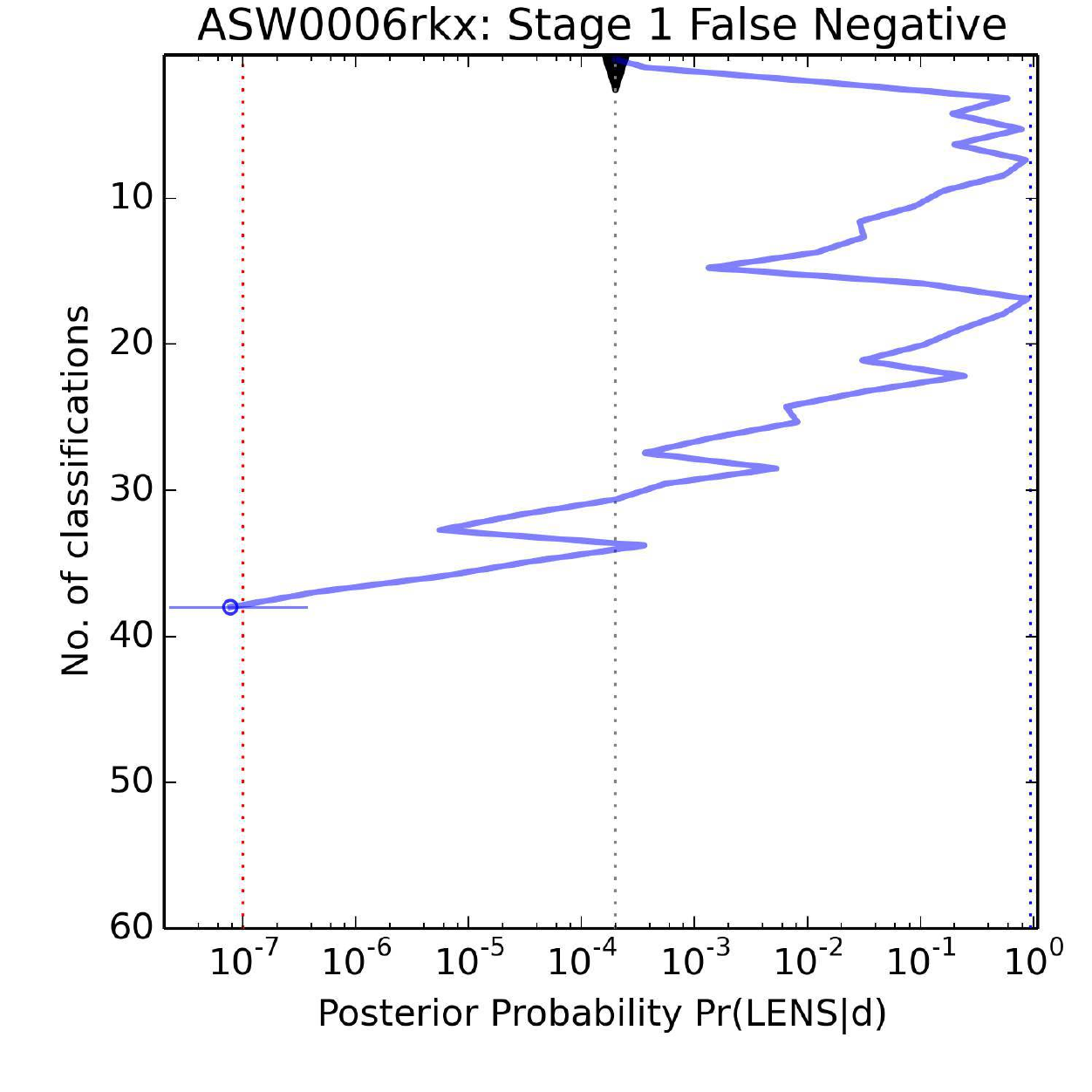}
    \end{minipage}
  \end{minipage}
\end{minipage}

\vspace{\baselineskip}

\begin{minipage}{\linewidth}
  \begin{minipage}[t]{0.47\linewidth}
    \begin{minipage}{0.46\linewidth}
      \centering\includegraphics[width=\linewidth]{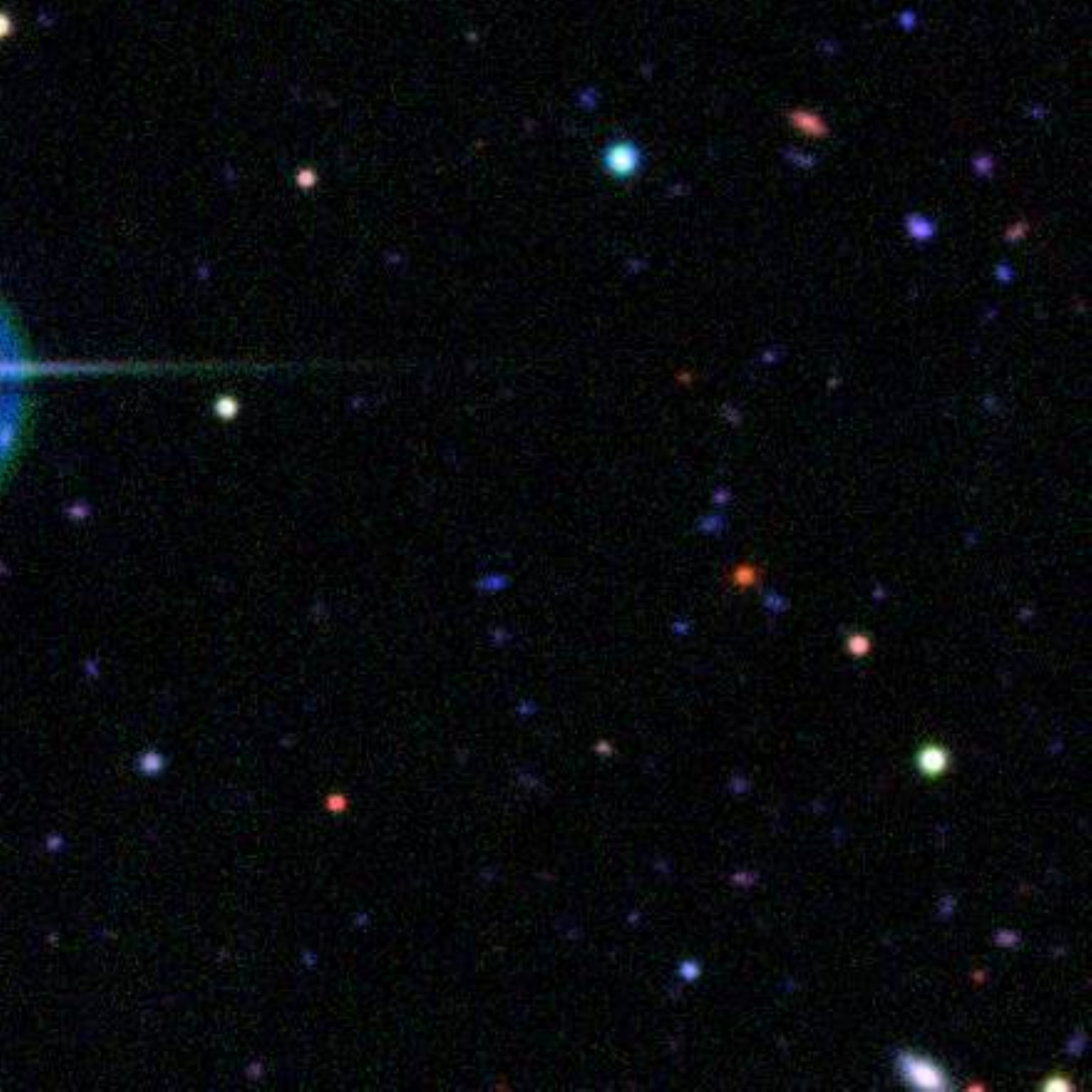}
    \end{minipage}\hfill
    \begin{minipage}{0.50\linewidth}
      \centering\includegraphics[width=\linewidth]{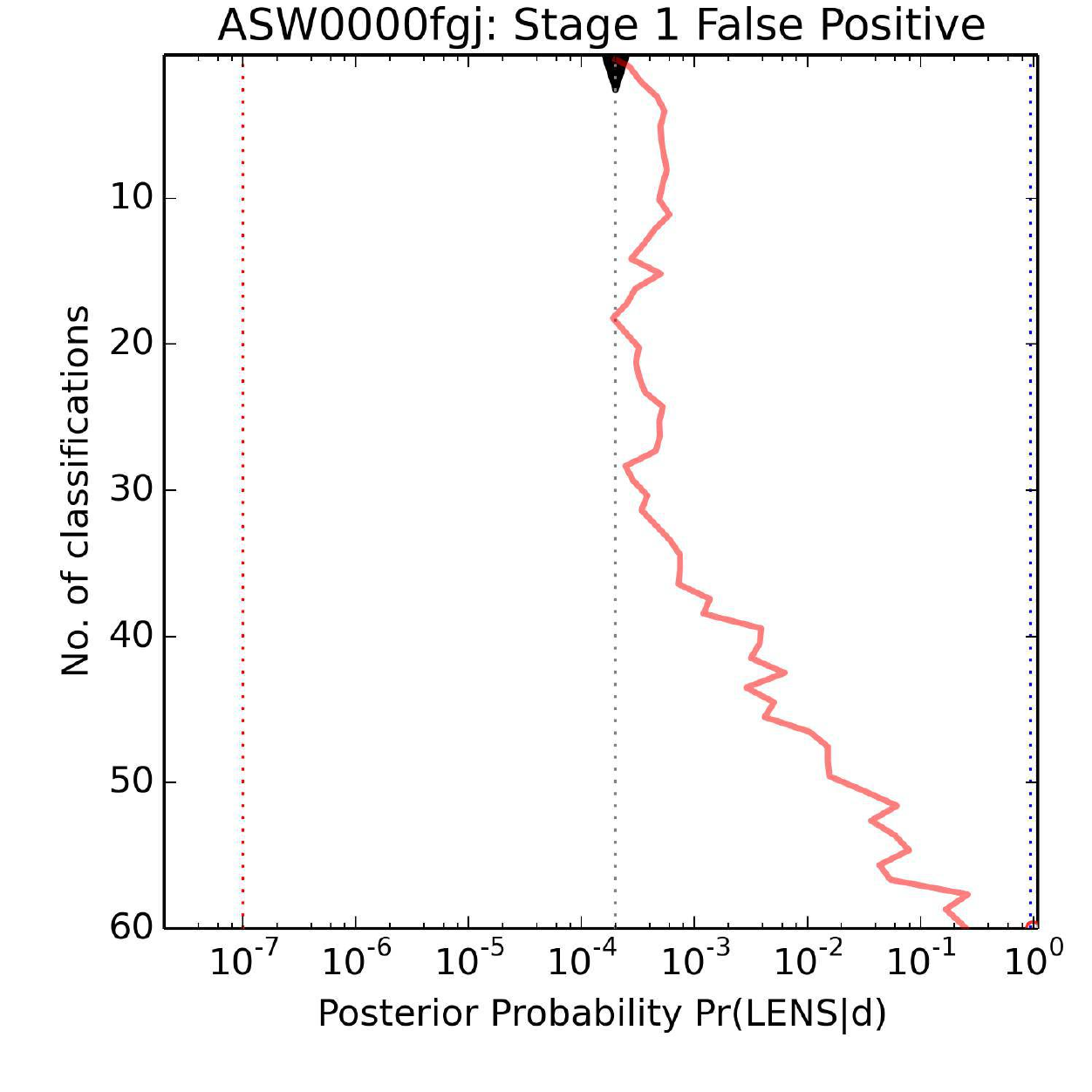}
    \end{minipage}
  \end{minipage}\hfill
  \begin{minipage}[t]{0.47\linewidth}
    \begin{minipage}{0.46\linewidth}
      \centering\includegraphics[width=\linewidth]{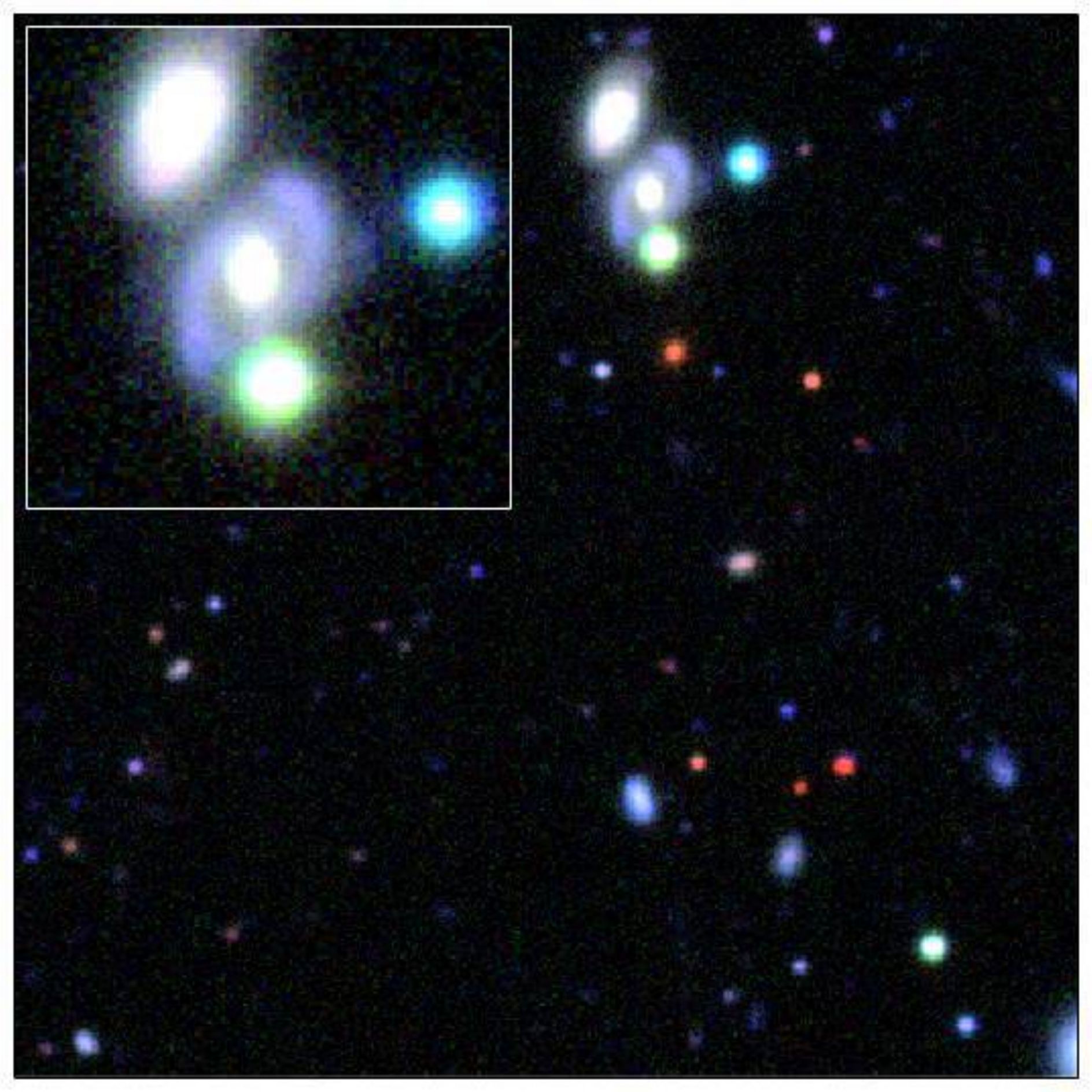}
    \end{minipage}\hfill
    \begin{minipage}{0.50\linewidth}
      \centering\includegraphics[width=\linewidth]{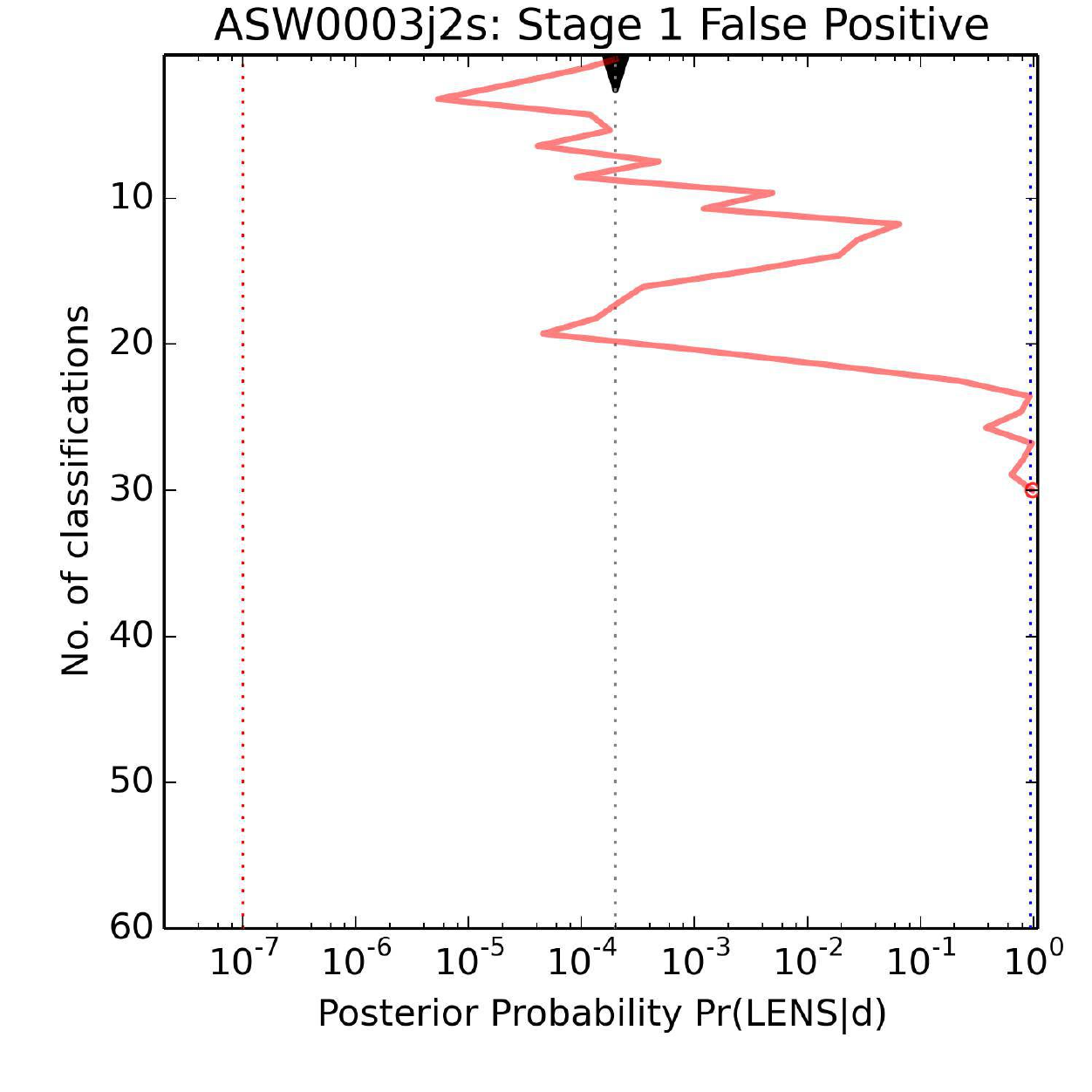}
    \end{minipage}
  \end{minipage}
\end{minipage}
\caption{Illustrative examples of
false negatives (blue) and false positives (red) from the classification of
the \SW$-$\cfhtls training set. The trajectory plot for each of the 4 subjects is
shown to the right of its image. Lefthand panels show the most common types of
trajectories: random walks with short steps resulting from no high skill classifiers
viewing the subject. The righthand panels show examples where higher skill
classifiers have been involved, causing larger, more efficient ``kicks''
(top) but also catastrophic mis-classifications (bottom right).
Note that both these
subjects were {\it very nearly} detected but didn't quite reach the threshold
set (marked by the blue dashed line). Insets for the false
negatives show the lens feature. The inset for the lower righthand panel shows
where most volunteers believed a lens was located. There is no corresponding inset
for the lower left panel because no particular feature stood out to the
citizens. Instead, the volunteers clicked many different features.}
\label{fig:discuss:performance:trajectories}
\end{figure*}

\subsection{Improving performance: reducing incompleteness and impurity}
\label{sec:discuss:performance}

We investigated the source of the incompleteness and impurity visible in
\Fref{fig:results:sample:CP}, by examining the \StageTwo ``false negatives''
(simulated lenses that incorrectly acquired $P < 10^{-7}$) and
false positives (duds or impostors that incorrectly acquired $P > 0.95$),
and their behaviour as they are classified using the online
analysis trajectory plots introduced in \Sref{sec:swap}
(\Fref{fig:swap:subject-trajectories}).
\Fref{fig:discuss:performance:trajectories} shows 2 example simulated lenses
that were missed ($\pr(\LENS|C,T) < 10^{-7}$) by the \SW system (top row),
and 2 example
non-lenses that were incorrectly flagged as candidates ($\pr(\LENS|C,T) >
0.95$) by the \SW system (bottom row).

In some cases the rejection of the false negatives is understandable: the
lensed features are faint, or in some cases, appear somewhat unrealistic
compared to real lenses.
However, in other cases a reasonably obvious lens was passed over.  This
mainly seems to be due to noise in the system: when only low-skill classifiers
view a subject, all the updates to its posterior probability are small, and if
none are very confident about the presence of lensing, the subject follows a
random walk down its trajectory plot. This can be seen in the top left panel
of \Fref{fig:discuss:performance:trajectories}.
The false positives show similar behaviour, for example in the bottom lefthand
panel of the figure.

As well as subjects being ``unlucky'' in this way, there
are two less common failure modes associated with mistakes made by higher skill
volunteers, illustrated in the righthand column of the figure. In the false negative
subject shown in the top right-hand panel, several high skill classifiers update
the subject upwards in log probability by some way each time, but other,
comparable skill classifiers mis-classify the system to lower probability.   The
trajectory looks like a random walk, but with bigger step sizes; this particular
subject came very close to crossing the detection  threshold three times, but
didn't quite make it. The bottom right-hand panel shows an example of a final,
apparently rare, failure mode: we see some short-step random walk behaviour,
followed by a mis-classification by a very high skill ``expert'' classifier
after 20 classifications that ``kicks'' the subject to high probability.

There are a number of places where we can address these problems and
improve system performance: adding flexibility to the classification
interface, educating the volunteers, assigning subjects for
classification, and interpreting the classification data.

Some of  the mistakes made by reasonably high skill classifiers working at high
speed could have perhaps been corrected by those classifiers themselves, had
they had access to a ``go back'' option. While clearly enabling error reduction,
we might worry about such a mechanism having a negative effect on
volunteer confidence: it may be that encouraging this sort of
checking would result in increased and not necessarily productive caution. This
could be tested by presenting a fraction of the volunteers with a version of the
site that actively suggested that they take this approach, and then tracking the
relative performance of ``collaborative'' classifications compared with
independent ones. We leave the exploration of this to further work.

Mistakes by both high and low skill classifiers could be reduced by improving
the training in the system, which could be done in several ways. One is to make
more training images available to those who want or need it. A basic level of
training images are needed for SWAP to build up an accurate picture of each
classifier's skill -- but one could imagine volunteers {\it choosing} to see
more training images (still at random) in their stream. We could also experiment
with providing greater training rates early on for all volunteers, although
this carries significant risk: retention rates may drop if too few ``fresh'' test
images are shown early on. Another way of improving the training could be to
provide more information about what gravitational lens systems look like. In
this project, the Spotter's Guide, and the Science, FAQ and About pages were
always available on the site, but as a passive background resource. We might
consider providing more links to this guide in the feedback messages shown to
the classifiers as they go -- or perhaps extending these messages to themselves
include more explanations and example images. We might also investigate a more
dynamic Spotter's Guide: a set of manipulable model lenses illustrating all the
possible image configurations that those deflectors can make could help
volunteers very quickly gain understanding of what lensed features can look
like. Such a tool is under development.\footnote{\texttt{http://slowe.github.io/LensToy}}

\begin{figure}
\centering\includegraphics[width=0.9\linewidth]{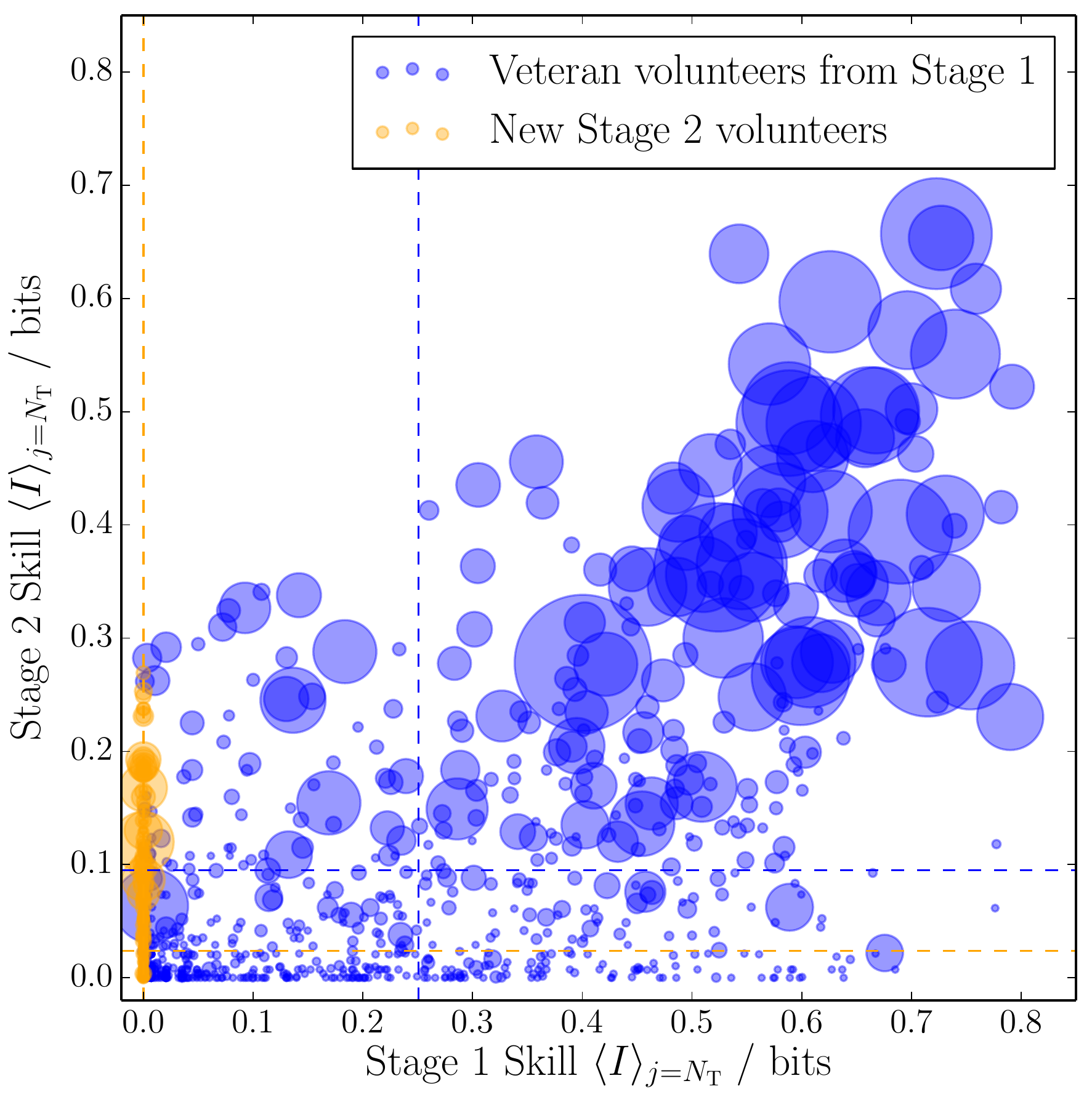}
\caption{\StageTwo classifier skill, as a function of their \StageOne skill.
Veterans from \StageOne  are shown in blue, while new volunteers are shown in
orange. Point size is proportional to the total contribution made,
while dashed lines are drawn at the mean values for each sample.}
\label{fig:discuss:performance:stage1vstage2}
\end{figure}

However, many false negatives (around 60\%, from inspection of a random sample)
seem to be due to the statistical noise associated with the many short,
semi-random direction kicks arising from classifications made by low skill,
inexperienced classifiers (\Sref{sec:results:crowd}). While improved training
will help reduce this effect, it is here that targeted task assignment could
also help improve system performance, by bringing higher skill volunteers in
when needed. We advertised \StageTwo of the current project to all registered
users; it was taken up by \StageOne veterans with a broad range of skill, and
also picked up a significant number of new users who did not have enough time to
acquire high skill (since \StageTwo was quite short): of the 1964 volunteers who
took part in the \StageTwo classification round, only 774 were veterans from
\StageOne. This issue is discussed further in \PaperTwo, where we investigate
the known lenses missed by the \SW system.

\Fref{fig:discuss:performance:stage1vstage2} shows how the \StageOne
skill maps on to \StageTwo skill and contribution. This figure suggests
that, while the gains are likely small (since  most of the contribution
is still made by high skill volunteers),  there could have been some
benefit to opening \StageTwo on the basis of \StageOne agent skill, in
order to reduce the noise in the system generated by new and low skill
volunteers in this more difficult classification stage.  In future,
dynamically allocating subjects to volunteers according to their agents'
skill could alleviate this issue. In particular one can imagine doing this
for subjects that have classification histories that indicate that they
are worthy of further study.
These are things we plan to
experiment with in future.

Finally, there could be some performance gains to be made by improving the SWAP
agent model, or its implementation.
Low skill is typically a result  of inexperience (\Sref{sec:results:crowd}) --
but it could be the agent that is inexperienced, as much the classifier.  In a
future paper we plan to investigate the use of the test images as well as the
training images in accelerating the agent's learning (Davis et al, in
preparation). We also plan to investigate the use of offline analysis at
\StageOne, partly for the same reason (see also \Sref{sec:discuss:efficiency}
below.) Given the above findings  about the effects of noise in the SWAP system,
we are also motivated to explore further the possible noise-reducing effects of
the doing the analysis offline (\Sref{sec:results:sample}).

One might also consider looking at
introducing more conditional dependences in the confusion matrix elements, to
allow for some classifiers having greater skill in spotting one type of lens
than another, or, more generally, as a function of lens property (such as
colour, brightness, and image separation). In the current model, all agents are
considered to be completely independent, whereas in fact we might expect there
to be significant clustering of the confusion matrix elements in the $\PD - \PL$
plane. A hierarchical
model for the crowd, with hyper-parameters describing the distribution of
confusion matrix elements across the population, may well accelerate the agent
learning process by including the notion of one agent being likely to be
similar to its neighbours in the parameter space. Finally, it is worth noting that the model of
\citet{IBCC} explicitly avoids the assumption of the agent confusion matrix
elements being constant in time (as was assumed in \Sref{sec:swap}),
allowing the development of volunteer skill
to be more accurately tracked -- and that they did see some time-evolution in
the supernova zoo classifiers' skill. Finding a way to incorporate such a
learning model into SWAP while retaining its online character
is an interesting challenge for future work.


\subsection{Improving efficiency}
\label{sec:discuss:efficiency}

\begin{table*}
\begin{center}
\caption{Total crowd and subject sample properties from the \cfhtls project.}
\label{tab:crowd:contributions}
\begin{tabular}{cccccccc}
  \hline
  \hline {Stage} & Subjects & Contribution                          & Agents & Skill      & Classifications          & Candidates & Information \\
                           & $\Ns$    & $\sum_k^{\Nv} \contribution_k$ (bits) & $\Nv$  & $\sum_k^{\Nv} \skill_k$ (bits) & $\sum_k^{\Nv} \thiseffort$ & $\Ncands$  & $\sum_j^{\Ns}\sum_k^{\Nv} \information_{j,k}$ (bits) \\
            \hline
                      1    & $427064$ & $1292016.3$ & $36982$ & $1471.9$ & $10802125$ & $3381$ & $91122.6$ \\
                      2    & $3679$   &   $21895.8$ &  $1964$ &  $102.4$ &   $224745$ &   $89$ &  $1640.4$ \\
  \hline \hline
\end{tabular}
\medskip\\
\end{center}
\end{table*}

\Tref{tab:crowd:contributions} shows the total effort, contribution, skill and
information generated in both \StageOne and \StageTwo of the \cfhtls project, with the
total numbers of agents and subjects for comparison.   These numbers allow us
to quantify the efficiency of the system.

The contribution per classification is defined in terms of a hypothetical
subject with lens probability of 0.5; one bit of information is needed to
update such a subject's lens probability to either zero or one. This means
that a maximally complete classification stage would yield a total
contribution (summed over all agents) equal to the number of subjects. The
ratio of this hypothetical optimum to the actual total contribution is
therefore a measure of the stage's inefficiency. We find our inefficiency (by
dividing column 2 by column 3) to be 33\% and 17\% in \StageOne and \StageTwo
respectively. In \StageOne, this inefficiency is due to the daily processing: we
were not able to retire subjects fast enough, and so they remained in the
system, being over-classified. Indeed, only 3705745 classifications were
needed to retire all the subjects: the ratio of this to the total number made
is 34\%. (The remaining 1\% is due to not all subjects being classified to 1
or 0 probability.) At \StageTwo, we did not retire any subjects at all; the
inefficiency in this case was by design, to give everyone a chance to
appreciate what they had found together. (An unwanted side effect of this
policy was noted in \Fref{fig:discuss:performance:stage1vstage2} in the
previous sub-section.)

It is clear that to increase the efficiency of the system we need to reduce the
time lag between the classification being made and its outcome being analyzed.
The optimal way to do this would be to have the web app itself analyze the
classifications in real time. This is under investigation for future projects.
There may still be a place for a daily or weekly offline analysis: this could
potentially reduce the false negative and false positive rates by
``resurrecting'' subjects that had been retired by the online system before the
software agents had time to learn enough about their classifier's high skill.

If, as expected, the system efficiency could indeed be improved by a factor of
three via real-time classification analysis, we would expect the characteristic
completion time $\tau_0$  in \Eref{eq:speed} to decrease by roughly the same
factor, suggesting that a dataset of $10^5$~images could be searched for lenses
by a crowd of $10^5$ volunteers in about 6 days.

The futuristic lens-finding problem sketched in \Sref{sec:intro}, where $10^5$
lenses are to be found in a $10^4$ square degree wide field imaging survey, is
approximately representative of the challenge facing both the LSST and Euclid
strong lensing science teams \citep{LSSTSciBook,EuclidSciBook}. What role could
citizen scientists at \SW play in lens searching in the next decade? From
\Eref{eq:speed}, and assuming $\tau_0 = 6$ days (as above), we see that for the
\SW crowd to be able to classify $10^8$ images of  photometrically-selected
massive galaxies and groups in approximately 60 days, it would need to contain
$10^7$ volunteers. This is only  a factor of ten larger than the current size of
the Zooniverse userbase.
Alternatively, suppose that automated  lens finding algorithms were able to
select as few as $10^5$ targets for visual inspection, corresponding to a sample
with 10\% purity and a surface density of $\sim10$ per square degree \citep[a
rate within reach of \RF, for example:][]{GavazziEtal2014}. Now
increasing $\tau_0$
by a factor of three to allow for more inspection time per subject,  we find
that in this case \SW could enable a crowd of just $10^5$ volunteers to assess
them in about three weeks.


\subsection{Increasing crowd capacity}
\label{sec:discuss:capacity}

Finally, we comment on the size of the first \SW crowd, and how the system might
be scaled up for future surveys. In \Sref{sec:results:crowd} and
\Tref{tab:crowd:contributions}, the following rough picture emerged for the \SW
crowd in this project: it consisted of a few $10^4$ volunteers, with a few
$10^3$ achieving considerable skill, and a few $10^2$ having the time to make a
significant contribution. Slightly more quantitatively, we might note that  the
total skill of the crowd, computed by summing the skill of all the agents, is a
measure of the effective crowd size, in the sense that a crowd of perfect
classifiers would be of this size. By this measure, the \StageOne crowd was
equivalent to a team of 1470 perfect classifiers, while the \StageTwo crowd was
equivalent to a team of 102 perfect classifiers. With this same crowd, we saw in
\Sref{sec:results:speed}  that surveys providing a few $10^4$ subjects would be
completed quickly, if the high contribution rate of the current crowd were to be
repeated.

There are (at least) two ways in which we might increase the numbers of
high-contribution volunteers for larger projects in future. The first is
simply to increase the total crowd size, and hope that a similar fraction of
volunteers make large contributions. Greater exposure of the website to the
public through mass media would help. Another option is to advertise the
project to new groups of volunteers by translating the website into other
languages (something which is now supported by the Zooniverse). A
multi-lingual userbase would come with its own set of challenges, especially
in terms of volunteers' continuing training and interactions on \Talk.

The second way to scale up the number of high contribution classifiers is to
increase the rate at which new volunteers become dedicated volunteers. Based on
feedback from the wider \SW community, this could potentially be achieved
through closer collaboration with the science team. It is also possible that
dynamically assigning subjects on the basis of the volunteer's skill could act
as an incentive to some volunteers to increase their contribution, although any
such approach would need to take into account the need to optimise not only for efficiency but also for the most interesting or pleasurable volunteer experience; a solution which gave every volunteer a very uniform experience, for example, is unlikely to succeed even if it appears optimally efficient.  Reducing the
rate at which new volunteers lose interest could also play a role. Anecdotally,
it seems fairly common for new volunteers to be wary of classifying at all, for
fear of introducing errors. Better explanation of how their early
classifications are analyzed could help assuage these fears:
\Sref{sec:results:crowd} shows that effectively down-weighting new volunteers'
classifications (by setting their agents' initial confusion matrix elements to
those of a random classifier) leads to best performance, a result which should
be of some comfort to the nervous volunteer.


\section{Conclusions}
\label{sec:conclude}

We have designed, implemented and tested a system for detecting new strong
gravitational lens candidates in wide field imaging surveys by crowd-sourced
visual inspection. The \SW web-based classification interface presents carefully
prepared colour composite sky images to volunteers, who mark features they
propose to  have been lensed. The participants receive ongoing training with a
mixture of simulated lens and known-to-be-empty images, and we use this
information to automate the interpretation of their markers. In our
first lens search we simply divided the \cfhtls imaging into some 430,000 tiles,
and collected over 11~million volunteer image classifications.

By analyzing the  classifications made of the training set, we conclude that
gravitational lens detection by crowd-sourced visual inspection works,  and in
the following specific ways:

\begin{itemize}

\item Participation levels were high (about 37,000 volunteers, contributing
classifications at rates between $10^3$ and $10^4$ images per hour), suggesting
that if this can be maintained, visual inspection of tens of thousands of images
could be performed in just a few weeks. An expanded crowd of $10^6$ volunteers
(which has already been reached in Zooniverse) would
be able to inspect a plausible sample of $10^6$ LSST or Euclid targets
on similar time scales.

\item Over the course of the two stages of the \cfhtls project, the
set of images was reduced by a factor of $10^3$ (from a few hundred thousand to
a few hundred), enough for an expert team to take over.

\item The ``skill'' of the volunteers correlates strongly with the number of
images they have classified.  Quantifying the ``contribution'' as the integrated
skill over the test subjects classified, we find that about 90\% of the
contribution was made by a few hundred volunteers ($1\%$ of the crowd) -- but
that the broad width of the skill distribution means that a few thousand volunteers
could have made comparable contributions had they classified more images.

\item The optimal true positive rate (completeness) and false positive rate in
the training set were estimated to be around $92-94\%$ and $<1\%$
(in both classification stages).
This FPR translates to a purity of approximately $15-30$\%. We find that even
higher purities were achievable at \StageTwo
(perhaps as a result of the least visible candidates already
being discarded): here, 100\% purity was reached at just under $90\%$
completeness. Because \SW is a supervised learning system, these numbers should
be taken to be upper limits to what we should expect for real lenses (which have
not been selected for a high visibility training set).

\item The simulated gravitational lenses that were missed were predominantly
galaxy-scale lenses with faint blue galaxy sources, whose lensed features are
difficult to distinguish from the light from the lens galaxy (consistent
with what we find also for real lenses, see \PaperTwo). We observed some
additional scatter, with some simulated lenses accumulating low probability
simply due to classification noise. Future searches will need to address
of these issues. We expect it to be especially interesting to try
crowd-sourcing the visual inspection of target systems which have
had their candidate lens galaxy light automatically subtracted off (as, for example, \RF
does), in a collaboration between humans and machines.

\end{itemize}

In \PaperTwo we present the results of this \cfhtls lens search in more detail,
in particular focusing on the detection of real lenses, and the comparison
between \SW and various automated lens finding methods. In this paper  we have
shown how \SW provides a lens candidate detection service, crowd-sourcing the
time-consuming work of visually inspecting astronomical images for
gravitationally-lensed features. We invite survey teams searching for lenses in
their wide-field imaging data to contact us if they would like our help.


\section*{Acknowledgements}

We thank Matthias Tecza, Stuart Lynn, Kelly Borden, Laura Whyte, Brooke Simmons, David Hogg,
Daniel Foreman-Mackey, Thomas Jennings, Layne Wright, Cecile Faure, Jonathan
Coles, Stuart Lowe, Alexander Fritz and Jean-Paul Kneib for many useful conversations about
citizen science and gravitational lens detection, and for helping guide the
discussion at \SW \Talk,
and to the Dark Energy Survey
and Pan-STARRS strong lensing science teams for their suggestions and
encouragement. We are grateful to the anonymous referee, whose comments
and suggestions helped improve the paper significantly.

PJM and ES also thank the Institute of Astronomy and Astrophysics, Academia Sinica
(ASIAA) and Taiwan's Ministry of Science and Technology (MOST) for their
financial support of the workshop ``Citizen Science in Astronomy'' in March
2014, at which some parts of the SWAP analysis was developed.

We thank all 36,982 members of the \sw community for their
contributions to the project so far. A complete list of registered
collaborators is provided at \texttt{http://spacewarps.org/\#/projects/CFHTLS}.
We also thank the anonymous referee for useful comments on the paper.

PJM was given support by the Royal Society, in the form of a research
fellowship, and by the U.S. Department of Energy under contract number DE-AC02-76SF00515.
AV acknowledges support from the Leverhulme Trust in the form of a research
fellowship.
The work of AM and SM was supported by World Premier International Research
Center Initiative (WPI Initiative), MEXT, Japan. AM acknowledges the support of
the Japan Society for Promotion of Science (JSPS) fellowship. The work of AM
was also supported in part by National Science Foundation Grant No.
PHYS-1066293 and the hospitality of the Aspen Center for Physics. 
%
%

The \sw project is open source.
The web app was developed at \texttt{https://github.com/Zooniverse/Lens-Zoo}, and was supported by a grant from the Alfred P. Sloan Foundation, 
while the SWAP analysis software was developed at
\texttt{https://github.com/drphilmarshall/SpaceWarps}.

The CFHTLS data used in this work are based on observations obtained with
MegaPrime/MegaCam, a joint project of CFHT and CEA/IRFU, at the
Canada-France-Hawaii Telescope (CFHT) which is operated by the National Research
Council (NRC) of Canada, the Institut National des Science de l'Univers of the
Centre National de la Recherche Scientifique (CNRS) of France, and the
University of Hawaii. This work is based in part on data products produced at
Terapix available at the Canadian Astronomy Data Centre as part of the
Canada-France-Hawaii Telescope Legacy Survey, a collaborative project of NRC and
CNRS.


\bibliographystyle{apj}
\bibliography{references_system}


\appendix


\section{Information Gain per Classification, Agent ``Skill'' and ``Contribution''}
\label{appendix:swap}

With an agent's confusion matrix in hand we can compute the
\emph{information} generated in any given classification. This will
depend on the confusion matrix elements (\Eref{eq:confmat}) but also on
the probability of the subject being classified containing a lens. The
quantity of interest is the relative entropy, or Kullback-Leiber
divergence, between the prior and
posterior probabilities for the possible truths $T$
given the submitted classification $C$:
\begin{align}
\information =& \sum_T \pr(T|C) \log_2 \frac{\pr(T|C)}{\pr(T)}     \notag \\
             =&    \pr(\LENS|C) \log_2 \frac{\pr(C|\LENS)}{\pr(C)} \notag \\
             +&    \pr(\NOT|C)  \log_2 \frac{\pr(C|\NOT)}{\pr(C)},
\end{align}
where, as above, $C$ can take the values $\saidLENS$ or $\saidNOT$.
Substituting for the posterior probabilities using \Eref{eq:app:first} we get
an expression that just depends on the elements of the
confusion matrix $\CM$ and the pre-classification subject lens
probability $\pr(\LENS) = p$:
\begin{align}
\information =    &     p \frac{\CM_{CL}}{p_c} \log_2 \frac{\CM_{CL}}{p_c}  \notag \\
                  & +(1-p)\frac{\CM_{CN}}{p_c} \log_2 \frac{\CM_{CN}}{p_c},
  \label{eq:app:infogain}
\end{align}
where the common denominator $p_c = p\CM_{CL} + (1-p)\CM_{CN}$. This
expression has many interesting features.  If $p$ is either zero or one,
$\information(C) = 0$  regardless of the value of $C$ or the values of
the confusion matrix elements: if we know the subject's status with
certainty, additional classifications supply no new information. If we
set $p$ to be the prior probability, \Eref{eq:app:infogain} tells us how
much information is generated by classifying it all the way to $p = 1$
(which a perfect classifier, with $\CM_{LL} = \CM_{NN} = 1$, can do in a
single classification). For a prior probability of $2\times 10^{-4}$,
12.3 bits are generated in such a ``detection.''  Conversely, only
0.0003 bits are generated during the rejection of a subject with the
same prior: we are already fairly sure that each subject does not
contain a lens! Imperfect classifiers (with $\CM_{LL}$ and $\CM_{NN}$
both less than 1)  generate less than these maximum amounts of
information each classification; the only classifiers that generate zero
information are those that have $\CM_{LL} = 1 - \CM_{NN}$ (or
equivalently, $\CM_{CL} = \CM_{CN}$ for all values of $C$). We might
label such classifiers as ``random'', since they are as likely to
classify a subject as a $\saidLENS$ no matter the true content of that
subject.

\Eref{eq:app:infogain} suggests a useful information theoretical
definition of the classifier skill  perceived by the agent. At a fixed
value of $p$, we can take the expectation value of the information gain
$\information$ over  the possible classifications that could be made:
\begin{align}
\langle\information\rangle   =& \sum_C \sum_T \pr(T|C) \pr(C) \log_2 \frac{\pr(T|C)}{\pr(T)} \notag \\
         =& - \sum_T \pr(T) \log_2 \pr(T) \notag \\
          & + \sum_C \pr(C) \sum_T \pr(T|C) \log_2 \pr(T|C) \notag \\
         =&         p  \left[ \mathcal{S}(\CM_{LL}) + \mathcal{S}(1-\CM_{LL}) \right] \notag \\
          &     +(1-p) \left[ \mathcal{S}(\CM_{NN}) + \mathcal{S}(1-\CM_{NN}) \right] \notag \\
          & - \mathcal{S}\left[ p    \CM_{LL}       + (1-p)(1-\CM_{NN})     \right] \notag \\
          & - \mathcal{S}\left[ p (1-\CM_{LL})      + (1-p)   \CM_{NN}      \right]
\end{align}
where $\mathcal{S}(x) = x \log_2{x}$. If we choose to
evaluate $\langle\information\rangle$ at $p = 0.5$, the result has some
pleasing properties. While random classifiers presented with  $p = 0.5$
subjects have $\skill = 0.0$  as expected, perfect classifiers appear to
the agents to have  $\skill = 1.0$. This suggests that  $\skill$, the
amount of information we expect to  gain when a classifier is presented
with a 50-50 subject, is a reasonable quantification of
\emph{normalised skill}. A consequence of this choice is that the
integrated skill (over all agents' histories) should come out to be
approximately
equal to the number of subjects in the survey, when the search is
``complete'' (and all subjects are fully classified). Therefore, a
particular agent's integrated skill is a reasonable
measure of that classifier's
\emph{contribution} to the lens search.

We conservatively initialize both elements of each  agent's confusion
matrix to be $\CM^0_{LL} = \CM^0_{NN} = 0.5$, that of a maximally ambivalent
random classifier, so
that all agents start with zero skill. While  this makes no allowance
for volunteers that actually do have previous experience of what
gravitational lenses look like, we might expect it to help mitigate
against false positives. Anyone who classifies more than one image (by
progressing beyond the tutorial) makes a non-zero information
contribution to the project.

The total information generated during the \cfhtls project is shown in
\Tref{tab:crowd:contributions}. Interpreting these numbers is not easy, but we
might do the following. Dividing this by the amount of information it takes to
classify a \SW subject all the way to the detection threshold (lens
probability 0.95), and then multiplying by the survey inefficiency gives us a
very rough estimate for the effective number of detections corresponding to
the crowd's contribution: these are 2830 and 25 bits for \StageOne and \StageTwo
respectively.  These figures are close to the numbers of detections given in
column 7 of the table.


\label{lastpage}
\bsp

\end{document}